\documentclass[useAMS,usenatbib,usegraphicx,dvips]{mn2e}
\usepackage{longtable}
\usepackage{color}
\newcommand{\et}{\sl et al. \rm}

\title[PS1 Nuclear Hypervariables]{Slow blue nuclear hypervariables in PanSTARRS-1}
 
\author[A.Lawrence et al]
{A.Lawrence$^1$, A.G.Bruce$^1$, C.MacLeod$^1$, S.Gezari$^2$, M.Elvis$^3$, M.Ward$^4$, 
\newauthor S.J.Smartt$^5$, K.W.Smith$^5$, D.Wright$^5$, M.Fraser $^{5,6}$,
\newauthor   P.Marshall$^7$
N.Kaiser$^8$, W.Burgett$^8$, E.Magnier$^8$, J.Tonry$^8$,K.Chambers$^8$, 
\newauthor R.Wainscoat$^8$,
C.Waters$^8$, P.Price$^9$, N.Metcalfe$^4$, 
S.Valenti$^5$, R.Kotak$^5$, 
\newauthor A.Mead$^1$,C.Inserra$^5$, T.W.Chen$^5$
A.Soderberg$^3$ \\
$^1$Institute for Astronomy, SUPA (Scottish Universities Physics Alliance), University of Edinburgh, \\ 
Royal Observatory, Blackford Hill, Edinburgh EH9 3HJ, UK \\
$^2$Department of Astronomy, University of Maryland, College Park, MD  20742-2421, USA\\
$^3$Harvard-Smithsonian Center for Astrophysics, 60 Garden St, Cambridge, MA 02138, USA\\
$^4$Department of Physics, Durham University, South Road, Durham DH1 3LE, UK\\
$^5$Astrophysics Research Centre , School of Mathematics and Physics, Queen's University Belfast, Belfast, BT7 1NN, UK\\
$^6$Institute of Astronomy, University of Cambridge, Madingley Road, Cambridge CB3 OHA, UK \\
$^7$KIPAC, SLAC National Accelerator Laboratory, 2575 Sand Hill Road, Menlo Park,
CA 94025, USA \\
$^8$Institute for Astronomy, University of Hawaii, 2680 Woodlawn Drive 
Honolulu, HI 96822, USA \\
$^9$ Department of Astrophysical Sciences, Princeton University, Princeton, 08544, USA.
}

\begin{document}

\date{Accepted 2016 August 04. Received 2016 July 29 ; in original form 2015 December 01 }

\pagerange{\pageref{firstpage}--\pageref{lastpage}} \pubyear{2013}

\maketitle

\label{firstpage}


\begin{abstract}
We discuss 76 large amplitude transients ($\Delta m>1.5$) occurring in the nuclei of galaxies, nearly all with no previously known Active Galactic Nucleus (AGN). They have been discovered as part of the Pan-STARRS1 (PS1) 3$\pi$ survey, by comparison with SDSS photometry a decade earlier, and then monitored with the Liverpool Telescope, and studied spectroscopically with the William Herschel Telescope (WHT). Based on colours, light curve shape, and spectra, these transients fall into four groups. A few are misclassified stars or objects of unknown type. Some are red/fast transients and are known or likely nuclear supernovae. A few are either radio sources or erratic variables and so likely blazars.  However the majority ($\sim 66\%$) are blue and evolve slowly, on a timescale of years. Spectroscopy shows them to be AGN at $z\sim 0.3 - 1.4$, which must have brightened since the SDSS photometry by around an order of magnitude. It is likely that these objects were in fact AGN a decade ago, but too weak to be recognised by SDSS; they could then be classed as ``hypervariable'' AGN. By searching the SDSS Stripe 82 quasar database, we find 15 similar objects. We discuss several possible explanations for these slow blue hypervariables - (i) unusually luminous tidal disruption events; (ii) extinction events; (iii) changes in accretion state; and (iv) large amplitude microlensing by stars in foreground galaxies. A mixture of explanations (iii) and (iv) seems most likely. Both hold promise of considerable new insight into the AGN phenomenon.

\end{abstract}



\parindent 0pt
\parskip 10pt

\section{Introduction} \label{sec:intro}

Searches for extreme optical extragalactic transients are of great interest in a variety of ways - as a method to find rare types of SNe, tidal disruption events around dormant black holes, rare blazars, and the possibility of accretion outbursts in Active Galactic Nuclei (AGN). In recent years systematic searches have been made using wide field instruments. These have been either targeted at finding SNe, or at finding candidates for Tidal Disruption Events (TDEs), typically with a fast rise and decay over months - for example with GALEX \citep{Gezari2008, Gezari2009}, SDSS \citep{VanVelzen2011}, PTF \citep{Cenko2012, Arcavi2014}, ASASSN \citep{Holoien2014}, and PanSTARRS-1 (PS1; \citet{Gezari2012, Chornock2014}). In addition the Time Domain Spectroscopic Survey (TDSS; \citet{Morganson2015}), a subset of the SDSS-IV programme, targets variable objects for follow-up spectroscopy, including already confirmed quasars that show more than 0.7 magnitudes of variability.

The PS1 nuclear transients reported so far \citep{Gezari2012, Chornock2014} have been based on data from the Medium Deep Survey, ten fields with 8 sq.deg. each, observed with a cadence of a few days. In this paper we report on a very large area search for large amplitude ($\Delta m > 1.5$ mag) nuclear changes in faint extragalactic objects, by comparing the PS1 $3\pi$ survey with the SDSS sky a decade earlier, over 11,663 sq.deg. Our original aim was to find candidates for TDEs, but in fact we seem to have found a class of slow-blue extreme AGN ``hypervariables'' at $z\sim 1$, with intriguing properties. These objects are statistically consistent with being an extrapolation of the extreme tail of more well known AGN variability, (e.g. \citet{MacLeod2012, Morganson2014}), but it is far from clear what the cause of the variability is, and whether it is the same as more normal AGN variability. We have collected spectra and carried out monitoring over the last few years, and find these objects to (mostly) show slow smooth order of magnitude outbursts over several years, to show large colour changes between the SDSS and PanSTARRS epochs, and to have weaker than average broad emission line strength. 

These transients were first reported in a conference paper by \cite{Lawrence2012a}. In this paper we present extensive results and analysis for these objects. In Section 2 we describe the Pan-STARRS1 programme, and the follow-up data taken with the Liverpool Telescope and the William Herschel Telescope. Section 3 presents basic data and analysis, including colours, light curves, and spectroscopic results. In Section 4 we present a further analysis, including luminosities, emission line properties, colour changes, and the statistics of variability in the context of AGN in general. In Section 5, we discuss four possible explanations of the cause of these slow smooth outbursts - TDEs, extinction events, accretion instabilities, and foreground microlensing.

\section{Observations} \label{sec:observations}

Our sample is based on Pan-STARRS1 and SDSS data, followed up with Liverpool Telescope (LT) monitoring, and spectroscopy with the William Herschel Telescope (WHT). We begin by describing each of these datasets, plus a small amount of additional data from other sources.

\subsection{The Pan-STARRS1 programme}\label{PS1}

Pan-STARRS1 (PS1) is a 1.8m optical telescope with a 7 square degree field of view, imaging onto a mosaic CCD camera with sixty detectors each with 4800$\times$4800 pixels of size 0.258\arcsec , operating at the summit of Haleakala on the island of Maui, Hawaii. The system is described more fully in \citet{Kaiser2010}. Images are obtained through a set of five filters designated $g_{P1}, r_{P1}, i_{P1}, z_{P1}, y_{P1}$, described in \citet{Stubbs2010} and \citet{Tonry2012}. Four of these are similar to the SDSS $g,r,i,z$ set. The fifth is a $y$-band filter covering roughly 0.92$\mu$m -- 1.05$\mu$m. The system was built by the University of Hawaii, but was operated by the PS1 Science Consortium (PS1SC : see http://ps1sc.org) up until March 2014. The telescope is now part of the ongoing Pan-STARRS2 project. The PS1 data will be publicly available through the MAST facility (https://archive.stsci.edu/)

PS1 undertook several surveys. The two major surveys, and the most important for extragalactic transients, are the Medium Deep Survey (MDS) and the 3$\pi$ survey, described in \citet{Magnier2013}. The MDS repeatedly imaged a set of ten individual PS1 fields, with a roughly four day cadence. The 3$\pi$ survey, as the name suggests, mapped three quarters of the sky. In any one filter the aim (subject to weather of course) was to visit each piece of sky four times per year. The filter-visits are spread out so that each piece of sky is visited twenty times a year in total. The 3$\pi$ survey began in May 2010, and completed in March 2014.  Table 1 shows the typical nightly depth, along with the predicted stacked depth after 3 years, compared to the SDSS survey depth in the equivalent filters. Note that this paper concerns only that part of the $3\pi$ survey that contains the SDSS region.

The PS1 images are processed by the PS1 Image Processing Pipeline (IPP), which performs a standard reduction sequence followed by object cataloguing, astrometry and photometry in the natural PS1 system. For the purposes of the current paper, $g_{P1}, r_{P1}, i_{P1}, z_{P1}, y_{P1}$ AB magnitudes are roughly equivalent to both the related SDSS magnitudes, and the Liverpool Telescope magnitudes (see next section). For the MDS, difference imaging is used to search for transient events; for the 3$\pi$ survey, which is the main focus of this paper, transients are located by comparing catalogue objects as described below. 

The catalogues produced by IPP were made available to the PS1SC on a nightly basis and ingested into a MySQL database at Queen's University Belfast. These were cross-matched with SDSS objects from the DR7 catalogue \citep{Abazajian2009}, looking for significant changes. These potential transients went through an extensive sequence of both automated and human filtering and quality control, as well as preliminary classification, described in more detail in \citet{Inserra2013}. This quality control process is intended to err on the side of reliability rather than completeness, which means that any derived population statistics are only approximate, as we discuss later. 

\begin{table}
\begin{tabular}{|c|c|c|c|c|c|}
 & \multicolumn{5}{|c|}{survey depths} \\
 Method & $g$ & $r$ & $i$ & $z$ & $y$ \\
\hline
3$\pi$(nightly, 5$\sigma$)& 22.0 & 21.6 &21.7 & 21.4 & 19.3 \\
3$\pi$(3 yrs, 50\%)  & 23.4 & 23.4 &23.2 & 22.4 & 21.3 \\
3$\pi$(est.final, pk cts)  & 23.0 & 22.8 &22.5 & 21.7 & 20.8 \\
SDSS (pk cts)           & 22.8 & 22.2 & 21.6 & 20.3 & --- \\
\end{tabular}
\caption{Characteristics of the 3$\pi$ survey. Row-1 gives the typical nightly 5$\sigma$ depth in AB magnitudes for the 3$\pi$ survey, estimated by \citet{Inserra2013}. The next two rows are from the Small Area Survey as analysed by \citet{Metcalfe2013}. Row-2 gives the estimated final depth of the 3$\pi$ survey, measured as the magnitude where counts are 50\% of their peak values, roughly equivalent to 5$\sigma$. Row-3 gives the magnitude at which source counts peak. Row-4 gives the depth of SDSS in the same piece of sky, estimated as the magnitude where source counts peak.} 
\end{table}

\subsection{SDSS data}\label{sec:SDSS}

The SDSS data we use in this paper comes from data release seven (DR7) \citep{Abazajian2009}). Although there have been subsequent SDSS releases, because selection was made from DR7, we have continued to use DR7 data for consistency. We have confirmed that the revised values from later releases make negligible difference. The magnitudes we have used for the pre-existing galaxies are the composite ``cmodel'' magnitudes, which use a linear combination of exponential and de Vaucouleurs light profiles, and should in general be the most appropriate estimate of total flux for galaxies, and which also agree with the PSF magnitude for stellar sources. We also use the standard template-fitting photometric redshifts calculated for DR7, as described in \citet{Abazajian2009} and on the SDSS web pages. In a few cases where they were available, we have also made use of SDSS-I, SDSS-II or SDSS-III (BOSS) spectra.

The SDSS observations are roughly a decade earlier than the PS1 observations. The majority of our objects come from the SDSS Legacy Survey, which began in 2000 and according to \citet{Abazajian2009} was essentially complete by July 2006. Around 10\% of our targets come from the SEGUE imaging stripes, which were observed during 2005-2008. This seems to under-represent the fractional area of SEGUE imaging in DR7 (28\%) which may be connected with the slow nature of most of our transients.

\subsection{Liverpool Telescope observations}\label{sec:LT}

Objects selected as nuclear transients as described below have been monitored with the Liverpool Telescope (LT). Although no new targets are being produced, the monitoring programme continues for existing targets. The LT observations give us denser sampling than provided by PS1, and also crucial $u$-band coverage. The LT is a 2.0m robotic telescope on the island of La Palma, operated by Liverpool John Moore's University. The system is described in \citet{Steele2004}. Observations from October 2011 onwards have used either the RATCAM or IO:O instruments, gradually converting to the latter.  RATCAM is a CCD camera with 2048$\times$2048 pixels of size of 0.135\arcsec , but normally used with 2$\times$2 binning. IO:O is a CCD camera with 4096$\times$4096 pixels of size of 0.15\arcsec , also normally used with 2$\times$2 binning. The field of view (4.6 arcmin for RATCAM and 10.2 arcmin for IO:O) provides many SDSS stars for photometric calibration, so that we do not have to rely on completely transparent conditions. Both systems have an extensive set of filters. We have used filters which closely approximate the Sloan filters. For the purposes of this paper, we take the derived magnitudes to be on the SDSS AB magnitude system, and designate the magnitudes simply $u,g,r$.  

 
The standard LT pipeline performs bias subtraction, flat fielding, and astrometric reduction before passing data files to users. We then measured target magnitudes using simple aperture photometry with a software aperture diameter of 2\arcsec , using SDSS DR7 catalogued stars in the field, of which there are typically several tens, as photometric calibrators. (LT makes occasional standard star observations during the night, but we did not use these). The seeing in our LT images varies between 0.7\arcsec and 2.0\arcsec. With seeing worse than this, we do not use the data. Most of our targets are dominated by the unresolved transient, so the aperture photometry is simple to interpret regardless of the seeing.  
 
Our usual strategy was to initially follow targets every few days or weekly, until it became clear how fast they were fading. Fast fading objects were followed until they were too faint to measure in a reasonable exposure time. More slowly changing objects are then monitored roughly fortnightly or monthly while they are in season. Targets brighter than $g=20$ are exposed for 100s each in $g$ and $r$, and 400s in $u$. Given the (very blue) colours of most of our targets, this gives 5\% photometry or better in all bands. For targets fainter than $g=20$ we use 200s in $g$ and $r$, but still use 400s in $u$ as attempting to maintain accurate $u$-band photometry becomes too expensive. 

\subsection{Sample definition for this paper}\label{sec:defn}

We have used a combination of PS1, SDSS, and LT data to construct our sample for further study.
The starting point is the ``Faint Galaxy Supernova Search (FGSS)'' programme run by QUB, as described in Inserra et al (2013). This starts with catalogued objects from nightly visits of the PS1 3$\pi$ survey in the SDSS footprint  (11,667 sq.deg. in DR7)  and cross matches with SDSS DR7 objects. Selection in any one filter requires that the PS1 object has a magnitude fainter than 15 and brighter than 20, and is within 3\arcsec of an SDSS object with magnitude between 18 and 23. To be selected as a transient, the change in magnitude between SDSS and PS1 has to be at least 1.5 mags in at least one of $g,r,i,z$ as compared to the respective matching filter. 

\citet{Finkbeiner2015} show that PS1 and SDSS photometric systems are consistent in these bands to $\pm 9$mmag, and much better after plate-to-plate adjustment of SDSS, down to at least $r=20$. Most of our objects are brighter than this in PS1, and fainter than this in SDSS, but systematic differences will be small enough that that a 1.5 mag difference is an extremely significant flux difference.  Likewise there are colour terms \citet{Tonry2012}, but they are small enough to be unimportant in transient selection. Of course, many of the SDSS magnitudes are of relatively low signal-to-noise, so that the precise interpretation of the flux difference is not so clear. This makes essentially no difference to the reality of transient detection (as is obvious from direct comparison of images), but is one of several reasons why we do not consider our list to be a statistical sample.

This selection routinely produced several thousand apparent transients per month. After both automated and eyeball quality control, this was reduced to around a hundred good transients per month. Most of the rejected objects are simply artefacts of one kind or another (see \cite{Inserra2013}), but some will be real transients, which makes it hard to construct reliable population statistics.

The 3\arcsec\ limit was aimed at finding supernovae. To select and study nuclear transients, we additionally required that the PS1 object was within 0.5\arcsec\ of an object previously classified morphologically as a galaxy in SDSS DR7. At $z=0.3$ in a standard cosmology this angular size corresponds to a linear scale of $\sim2$kpc. In addition, as a comparison, we included in our follow-up two objects morphologically classified as stars in DR7. Of these, one (J061829) turned out to be a cataclysmic variable; the other (J083544) turned out to be a quasar at $z=1.327$. Selection of such nuclear transient candidates began in Oct 2011, but was not done systematically until mid-2012. The selected objects were then monitored with the LT, as described in section \ref{sec:LT} above. The chain of selection concentrates on reliability rather than completeness. This means that our sample will give only a lower limit for event rates. However it should be representative of the properties of nuclear transients. It should also be noted that because our baseline is SDSS, from a decade before PS1, we are sensitive to long term changes, as opposed to a season-by-season comparison within the PS1 data, which would only be sensitive to short-term changes.

For this paper, we wanted to have a reasonably long stretch of follow-up coverage with LT. We have therefore defined the sample for present study as those selected as described above, that had at least three LT photometry epochs by May 2013. (Most have many more photometry points since.) Table A1 lists the 76 targets that meet these criteria, with some basic information.

The requirement for the SDSS object to be classified as a galaxy was made because originally we were hunting for Tidal Disruption Events (TDEs). However as we shall see, most of the objects we have found are in fact AGN that were presumably just below the detection threshold in SDSS imaging (see section 4.1). There may then be objects already classified as AGN which are just extremely variable. This is in fact the case, as we show in section 4.5.

\subsection{WHT spectroscopy}\label{sec:WHT}

Since late 2012 we have been collecting spectroscopic observations of our targets with the Intermediate dispersion Spectrograph and Imaging System (ISIS) on the William Herschel Telescope (WHT). The WHT is a 4.2m telescope on the island of La Palma, Spain, and is part of the Isaac Newton Group of Telescope (ING). ISIS is a high-efficiency, double-armed, medium-resolution spectrograph. 

The WHT observations were made with the ISIS spectrograph using the standard 5300\AA\ dichroic and the R158R/R300B gratings in the red/blue channels.  This gave a spectral resolution of ~1500 at 5200\AA\ in the blue arm and ~1000 at 7200\AA\ in the red arm for a typical slit width of 1".  The GG495 order sorting filter was used in the red channel and both detectors used 2x binning in the spatial direction.  Reductions were performed using custom PyRAF scripts and the mean extinction curve for the observatory was assumed when performing the flux calibrations.

Table A1 lists the observation dates at the WHT. We have more than one epoch for a number of objects, but the date given corresponds to the data we describe and analyse later in this paper, and is generally near the peak of the light curve. In this paper we include results from all spectra taken by December 2014 by which time we had  spectroscopic data for 51/76 (66\%) of our sample (46 of these are new spectra, nearly all from WHT). A small number of objects have spectroscopic information from other sources - in particular from the INT, from the NOT telescope, and from the PESSTO programme on the ESO NTT telescope. 

Most of the spectra were taken under photometric conditions, but as ever with spectroscopy, seeing changes and centring issues mean that absolute photometry is probably reliable only to 20\% or so. (The relative spectrophotometry is much more accurate.) Some sessions had thin cloud. We have used a smooth interpolation between Liverpool Telescope $g$-band photometry points to calibrate these spectra. Given the short timescale variability seen on top of the long term trends that that we will discuss in section \ref{sec:results}, this calibration is also likely to be accurate to around 20\%. 

\subsection{Other data}\label{sec:otherdata}

In Table A2 we show associations with sources in other relevant surveys. 

(i) By definition, our targets are also objects in SDSS from DR7 \citep{Abazajian2009}. Table A2 shows the standard IAU designation for these objects. (ii) A large fraction are also detected in one or other of the UKIDSS surveys \citep{Lawrence2007}. The identifier here is again the standard IAU positional designation. (iii) Sixteen of our targets have also been detected as transients by the Catalina Real Time Transient Survey (CRTS; \citet{Drake2009}). The identifiers here follow the nomenclature from the CRTS website, specifying the telescope concerned, the trigger date, and the position of the object. (iv) Finally, we have searched the combined radio catalogue of \citet{Kimball2014} for sources withinn 30\arcsec\ of our targets, finding six objects. In Table A2 we show the sequential source number from the NRAO VLA Sky Survey (NVSS; \citet{Condon1998}). Most of these sources are also detected in FIRST, GB6, WENS, or VLSS.

\section{Results}\label{sec:results}

\subsection{Colours and amplitudes}\label{sec:colours}


\begin{figure}
\centering
\includegraphics[width=0.35\textwidth,angle=-90]{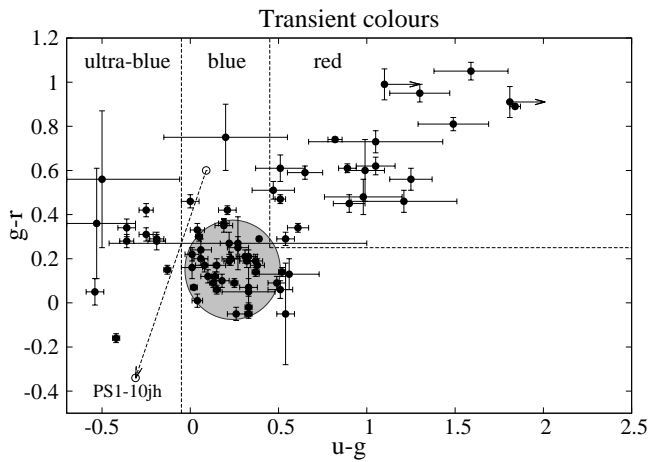}

\caption{\it\small Colours of the PS1 nuclear transients. The horizontal and vertical dashed lines divide the plane into colour classes, as explained in the text. The grey translucent ellipse shows the location of 90\% of SDSS spectroscopic quasars (see text). The points connected by arrows shows two versions of PS1-10jh, the TDE candidate from \citet{Gezari2012}. The upper circle is the version published by Gezari et al, based on difference imaging; the lower circle is the LT 2\arcsec\ photometry version, which includes the host galaxy contribution.}
\label{fig:trans-col}
\end{figure}


Table A3 shows the basic photometry results - the SDSS photometry, and the LT photometry near the time that the transient was first flagged by the PS1-QUB system  - typically an average of the first three--five LT epochs. (Note that no Milky Way extinction correction has been applied, but this is always small.)

Fig \ref{fig:trans-col} shows the $u-g$ vs $g-r$ colours of the transient objects. There is a reasonably clear distinction between red and blue objects. For future reference, we define the red objects as those with $u-g>0.45$ and $g-r>0.25$. The majority of the blue objects are consistent with the colours of quasars at moderate redshift  (e.g. \citet{Richards2001}). Taking the 
SDSS quasar catalog of \citet{Shen2011} we have extracted a reduced catalogue of 31,502 quasars with photometric errors less than 0.03 magnitudes on all of $u$, $g$, and $r$. The grey ellipse shows a colour range including 90\% of these quasars. A significant fraction (14\%) of our blue objects are much bluer than the typical quasar. For future reference we define ultra-blue objects as those with $u-g<-0.05$. These colour classifications - red, blue, and ultra-blue - are listed in Table A4.

In our reduced SDSS catalogue, only 1091 (3.5\%) are as blue as our ultra-blue objects, compared to 11 of our sample (14\%). This seems to show a significant over-representation of such ultra-blue objects in our sample. However, as we discuss in section \ref{sec:spectra}, we believe that almost all of these objects are AGN that happen to have strong line contamination near a relevant band-centre. The fraction presumably differs from SDSS overall because of the specific redshift distribution, which isn't the same as the SDSS quasar sample. In many of the figures that follow, we separate the colour classes, but in all the diagrams we explored, there was never any significant difference between the blue and ultra-blue objects, so for simplicity we do not distinguish them in the figures that follow.

\begin{figure}
\centering
\includegraphics[width=0.35\textwidth,angle=-90]{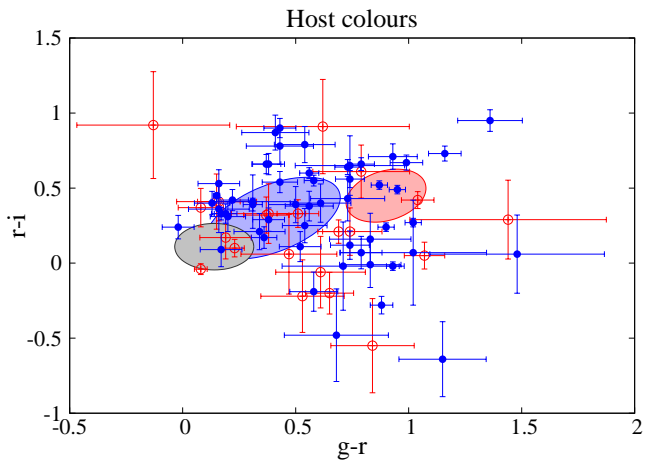}

\caption{\it\small Colours of the pre-existing SDSS host galaxies, from the SDSS DR7 photometry. The objects are divided by the colour of the transient object, defined as in the text and Fig. 1, with blue (filled) circles representing the blue and ultra-blue transients, and red (open) circles representing the red transients.
The light grey ellipse represents typical quasar colours, as in Fig. 1. The blue (middle) ellipse and the red (right hand) ellipse indicate the location of blue cloud and red sequence galaxies respectively, out to $z=0.22$, from \citet{Blanton2003a}.}
\label{fig:host-col}
\end{figure}

Fig \ref{fig:host-col} shows the $g-r$ vs $r-i$ colours of the pre-existing SDSS host galaxies. (The $u-g$ colours are too noisy to be informative on such a plot). For objects where the transient is classified as blue or ultra-blue (see Fig. 1), the host galaxies are almost always redder than the transient, with the median $\Delta (g-r) \sim 0.4$. (Colour changes are discussed in more detail in section \ref{sec:colour}.) 
The host colours show a rather large spread. At $g\sim 22$ we might expect galaxies to be at redshift $z\sim 0.2$. Fig. \ref{fig:host-col} shows  representative colours from \citet{Blanton2003a} for galaxies between $z=0$ and $z=0.22$, with a range of types. Many of the galaxies are consistent with this range,  but a few have relatively blue colours that could be consistent with either being AGN dominated, or being very late-type (star forming) galaxies. A number have rather peculiar colours, being for example blue in $r-i$ and red in $g-r$. This could be partly due to emission lines in cases of strong AGN contamination, or a mixture of an AGN with a $z\sim 1$ red host. 


\begin{figure}
\centering
\includegraphics[width=0.35\textwidth,angle=-90]{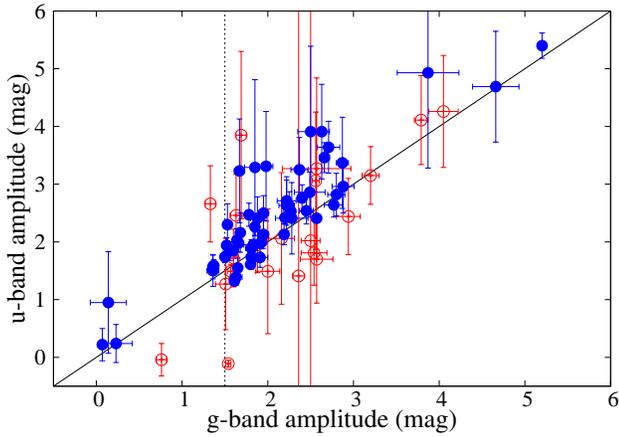}

\caption{\it\small Amplitudes of the transients in $u$ and $g$, divided by colour-class as defined in the text. Red (open) circles are the red objects, blue (filled) circles the blue and ultra blue ones. The diagonal line shows equality. The vertical dashed line shows the nominal 1.5 mag trigger level. These amplitudes are lower limits, as discussed in the text.}
\label{fig:amp-amp}
\end{figure}

Table A4 lists the $g$-band amplitudes of the transients, along with other information we will discuss later. The amplitude is the PS1 magnitude at the time of flagging, minus the SDSS DR7 magnitude. Fig. \ref{fig:amp-amp} compares the transient amplitudes $\Delta g$ and $\Delta u$. The median $\Delta g$ amplitude is 1.94 mag (a factor 6), but of course  these are lower limits, as we do not know how far below the SDSS flux the transient component was at the time of the SDSS measurement. For the red objects, $\Delta u$ is normally smaller than $\Delta g$, whereas for the blue and ultra-blue objects it is almost always larger, and within the errors could be larger in all cases. (The median $\Delta u$ is 1.81 mag for the red objects and 2.41 mag for the blue and ultra-blue objects.) This does not necessarily mean that the transient itself has a larger amplitude in $u$ than $g$ - almost certainly it simply reflects the fact the host galaxy is redder than the transient, so the contrast is stronger in $u$. 

Fig. \ref{fig:amp-amp} also shows thats some objects fall below the normal 1.5 mag trigger level for our study.
This occurs because the flagging was in the $r$, $i$ or $z$ bands, with the first $g$ band observation being a little later. For most objects this makes little difference, but in some cases the $g$ band or $u$ band flux had already fallen below the nominal trigger level - sometimes marginally so, sometime strongly so. 

In most cases, our observed amplitudes are large enough that the underlying galaxy will have only a small effect on the transient colours. However this may not always be the case. This can be illustrated by the case of PS1-10jh, the TDE candidate reported by \citet{Gezari2012}, which we include in some figures in this paper for comparison purposes, although it came from the MDS survey rather than the 3$\pi$/FGSS survey. (LT data were also taken). This means that its light curve data came from difference imaging and/or galaxy subtraction. These transient-only data are shown in Fig \ref{fig:trans-col} by an open circle. Shown as a solid circle is the simple aperture photometry point from LT data. It can be seen that using the galaxy-subtracted data changes $g-r$ by $\sim 1$ magnitude, transforming this object from blue (filled circle) to ultra-blue (open circle). However, such issues should affect only a small number of objects.


\subsection{Early decay}\label{sec:early-decay}


\begin{figure}
\centering
\includegraphics[width=0.35\textwidth,angle=-90]{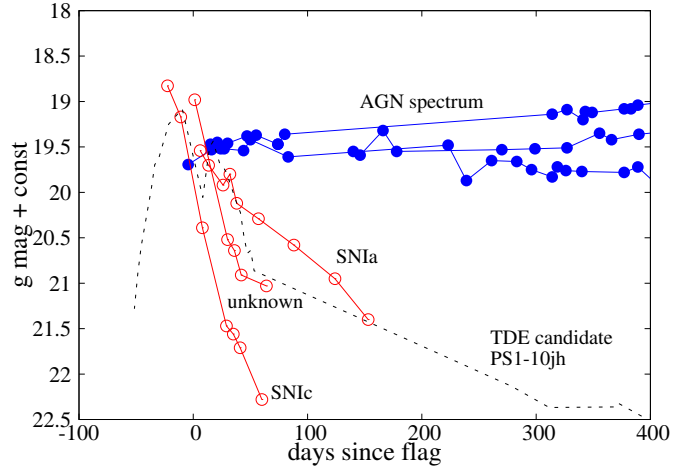}

\caption{\it\small A selection of light curves showing the differences between objects apparent during the first few months of monitoring. A constant (typically of plus or minus a few tenths of a magnitude) is added to the data for each object, to aid the clarity of the illustration. The symbols represent the colour of the transients (red,  and blue/ultra-blue) as in Fig. 3. Note that the flag date does not necessarily represent the peak of the light curve. The data for J160928=PS1-10jh are taken from \citet{Gezari2012}. For spectroscopic information, see Section \ref{sec:spectra}. }
\label{fig:lc-1yr}
\end{figure}



\begin{figure}
\centering
\includegraphics[width=0.35\textwidth,angle=-90]{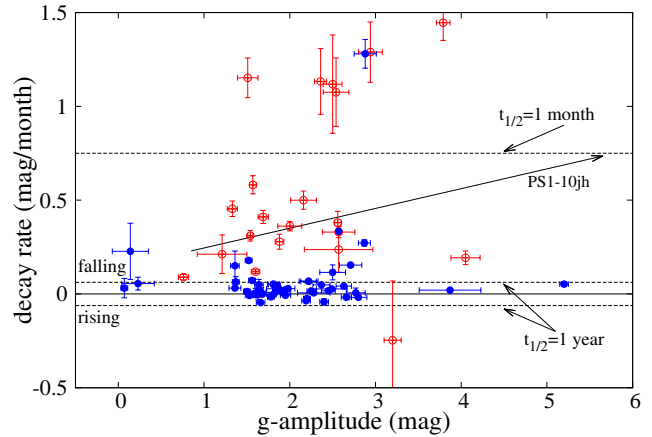}

\caption{\it\small Initial decay rate versus transient amplitude. Colour classes are indicated by symbol - open circles are red, filled circles are blue and ultra-blue. Note that J012514 is outside the plot - it is at $\Delta g=4.66$, slope=4.43 mag/month. The horizontal lines show fixed values of decay timescale, calculated by converting from decay rate in mags/month to the corresponding timescale on which flux falls by a factor two. The data for the TDE candidate J160928=PS1-10jh are taken from \citet{Gezari2012}. The two versions, connected by an arrow, show the results measured including the background galaxy (lower left) and after difference imaging (upper right).}
\label{fig:amp-slope}
\end{figure}


From the earliest monitoring, a clear distinction was apparent within the sample - red objects decayed fast, and blue objects decayed slowly, or were even consistent with being still rising. Where the spectral type was known, the fast-red objects were SNe and the slow-blue objects were AGN.
 This is illustrated by examples in Fig. \ref{fig:lc-1yr}. We quantified these effects by characterising each light curve by a simple linear slope, in magnitudes per month, estimated over the first three months if this slope was clearly changing. (Here ``month'' is taken as 30 days.) Some of the faster decaying objects were in fact too faint to measure before the three months was up. Note that falling by a factor of two in three months, i.e. so that $t_{1/2}=3$ months, means that the slope is 0.25 mag/month. Table A4 compiles the early-slope results, along with the transient amplitudes and colour classes from section \ref{sec:colours}, and the spectral classification from section \ref{sec:spectra}. In Table A4 we also show the decay corrected for time dilation, on the assumption that the causes are intrinsic to the source at the redshift found. 
 
In Fig. \ref{fig:amp-slope} we compare the derived decay rates and $\Delta g$ amplitudes, divided by colour class. The red objects nearly always show slopes of 0.2--1.5 mag/month, corresponding to fluxes with a two-folding timescale of weeks to months. The blue and ultra-blue objects have a median slope of $0.03$ mag/month, corresponding to fluxes with a two folding timescale of 2 years. Most of the blue and ultra-bue objects are falling, but some were apparently rising during this early period, and some still are. As we discuss later, the blue and ultra-blue objects are AGN; the time dilation correction makes their decay rates faster by factors ranging from 1.5 to 2.5, but the difference with the red objects is still very clear.

With some investigation, the distinction between light curve types is even clearer. Five blue or ultra-blue objects have slopes larger than 0.2 mag/month. One of these (J172639) has a large error bar on the decay rate and is consistent with being flat. 
J111706 is the solitary blue point with very large decay rate in Fig. 5, but is has a very large error on $u-g$ and so its classification as a blue object is unsafe. Likewise, J154950, which is classified as blue but which is in fact a supernova (see section \ref{sec:spectra}) also has a large $u-g$ error and so is probably not really blue. Next, J142446, although it has a blue $g-r=0.14$, also has a red $u-g$=0.52 and so only just fell within our blue classification. It also turns out to be a supernova (see section \ref{sec:spectra}). The only clear exception to the red-blue divide is J012514, which turns out to be an emission line star (see section \ref{sec:individuals}). 

We also show the TDE candidate PS1-10jh \citep{Gezari2012} for comparison. Once again it is quite distinct, with a large amplitude ($\Delta g =5.85$)  and a rapid decay (slope = 0.85 mag/month).

\subsection{Three year light curves}\label{sec:threeyear-lc}

Appendix B shows the full $\sim 3$ year PS1 + LT light curves in the g-band up to December 2014 for all the sample objects. Examples are shown in Fig. 6. We have categorised the light curves according to light curve shape, colour type, and the spectroscopic classification of section \ref{sec:spectra}. The results are summarised in Table 2. Below we discuss this categorisation.

Eight objects are known supernovae, from our spectroscopy or elsewhere, and they are all red and decay rapidly. A further eight objects have no spectra but are also red and decay rapidly, and so are presumed supernovae. 

Six objects are known radio sources. Of these, four were found by us to be AGN, and so the other two very likely are also AGN. Several of these objects are erratically variable, as blazars are known to be. One (J094309) is smoothly evolving, like the AGN we will discuss next; however we note that the radio match in this case is 28\arcsec\ and so may be spurious (see Table A2).  

\begin{figure}
\centering
\includegraphics[width=0.35\textwidth,angle=-90]{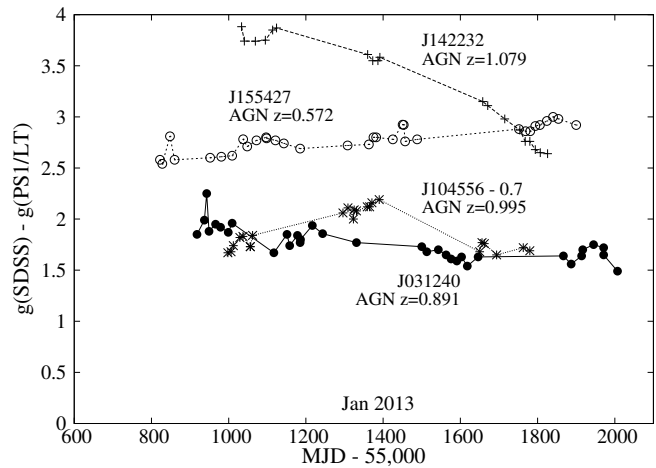}

\caption{\it\small Examples of slowly evolving AGN light curves over a three year period. The vertical axis is the difference between the magnitude at the date concerned, and the SDSS DR7 magnitude.}
\label{fig:lc-3yr}
\end{figure}

Excluding the radio sources, 35 objects are now known (all but one from our new spectroscopy) to be AGN. Of these, 16 are fading over a timescale of years, with a smooth decline that seems quite different from typical AGN, as well as being of larger amplitude, as we will discuss later. Twelve objects are rising, or have risen to a peak and begun falling again, but like the falling AGN, doing so in a fairly smooth manner. In Fig. \ref{fig:lc-3yr} we show examples of smoothly evolving AGN with falling, rising, and peaked light curves.


Seven further AGN don't quite fit the smoothly-falling-or-rising description. We have categorised them as ``complex'', but mostly they look as if they have a second peak.
Eight more objects do not yet have spectra, but are likely to be AGN - they are slowly evolving, and with two exceptions, blue. Of the two red objects in this category, one is falling slowly, and the other is falling relatively fast, but not as fast as the typical supernova. 

Of the remainder, for four objects the light curve data quality was too poor to say anything sensible; four turned out to be emission line stars, and two look likely to be supernovae. The one interesting remaining object, J133155, is blue and has a intermediate decay rate.

\begin{table*}
\begin{tabular}{|l|r|c|l|}
 Category & Number & Figure & Notes  \\
\hline
Known and probable SNe & 16 & B1,B2 & red, fast \\ 
Radio sources & 6 & B3 & 3 known AGN; most erratic \\ 
AGN, falling & 16 & B4,B5 & smooth, slow \\ 
AGN, rising & 7 & B6 & smooth, slow \\ 
AGN, peaked & 5 & B7 & smooth, slow \\ 
AGN, complex & 7 & B8 & mostly with two peaks \\ 
probable AGN & 8 & B9 & mostly smooth, slow, blue \\ 
emission line stars & 4 & B10 & blue, fast or erratic \\ 
unknown type & 7 & B10,B11 & mostly poor light curve quality \\ 
\end{tabular}
\caption{Classification of three year light curves. Light curves for all objects shown in Appendix B. Examples shown in Fig. 6}
\end{table*}

\subsection{Decade long light curves}\label{sec:tenyear-lc}

Sixteen of our objects were also triggered as transients by the Catalina Real Time Transient Survey (CRTS; \citet{Drake2009}). The CRTS identifications are listed in Table A2. For these objects, we were able to extract the retrospective CRTS data from their public data release, and, together with the SDSS magnitudes, make light curves which are from ten to thirteen years long. These light curves are shown in Figs B12-B13. The CRTS data points need to be treated with caution, both because they are typically of fairly low signal-to-noise (we have used multiple-epoch averaging in several places), but also because, while calibrated to the Johnson $V$-band, the CRTS data was taken with a white light filter which will have significant colour effects. Although we need to be quantitatively cautious, qualitatively the general pattern is clear.

Eight of these sixteen objects are known or likely supernovae based on our analysis so far. All of these show no previous history of variability - their light curves are flat, followed by a sharp rise and a decay over months, exactly as expected for a supernova event, although potentially some could TDEs.

\begin{figure}
\centering
\includegraphics[width=0.35\textwidth,angle=-90]{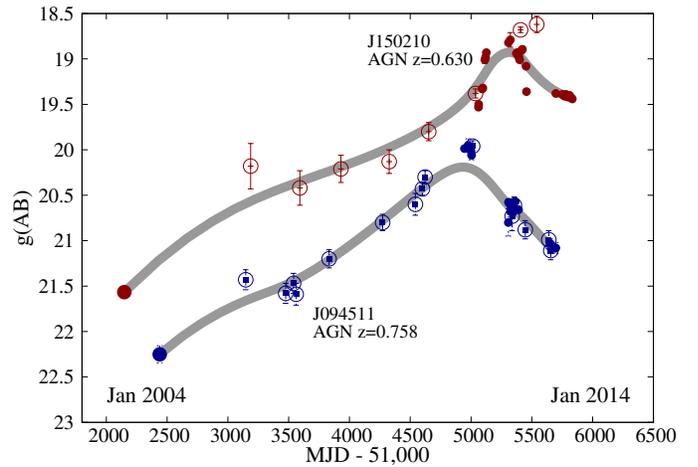}

\caption{\it\small Examples of slowly evolving AGN light curves over a ten year period, coming to a clear peak and then declining. The open circles represent CRTS data, seasonally averaged. (CRTS data with less binning is shown in Figs B12-B13) The filled circles to the right represent the PS1 and LT data; the filled circles to the left represent the SDSS era photometry. The grey curves are smooth polynomial Bezier curves simply meant to guide the eye.
}
\label{fig:lc-10yr}
\end{figure}

Four objects were found by us spectroscopically to be AGN. These all show a slow smooth rise leading up to the PS1 detection. They have all peaked and are now declining. The two clearest examples are shown in Fig. \ref{fig:lc-10yr}. The most interesting is J150210 which seems to show inflections in its light curve, at MJD=56,000 and 56,500.

Of the remaining four objects with long term data, one (J121834) is a radio source, and likely to be a blazar.   One is an object we found spectroscopically to be an emission line star. Two others are of uncertain nature, but could be supernovae.

For the radio sources, we have also looked for signs of variability by comparing fluxes from different surveys (using the combined catalogue of \citet{Kimball2014}). J094309 differs by two orders of magnitude between NVSS and VLSS, and J160329 by one order of magnitude. However in the latter case, FIRST and NVSS agree well. In other cases there is either no obvious sign of variability, or simply insufficient evidence, e.g. only seen in one survey.


\subsection{Spectroscopic results}\label{sec:spectra}

Table A4 summarises the known spectroscopic information for our sample, which is dominated by the new spectroscopy which we have collected. We collected spectra for 47 objects from the WHT as part of this programme; in addition we have spectral information available for three objects from the Nordic Optical Telescope (NOT; J094612), from the Isaac Newton Telescope (INT, J122417); and the Palomar 5m (P5m, J221441), all of which were collected as part of the related FGSS supernova programme. A further object (J105040) has a spectrum from SDSS, but no WHT spectrum - it was morphologically classified as a galaxy, but observed spectroscopically as a ROSAT target, and found to be an AGN at $z=0.306$. Finally we note that J081916 has both a WHT spectrum and an earlier SDSS spectrum, which was likewise obtained because it was a ROSAT target. Of these 51 objects with spectral information, 8 were SNe, 4 were variable stars, and 38 were AGN. The remaining object (J025633) had two spectra near peak (from WHT and NOT) which were very blue and featureless. J025633 is a radio source, and so is likely to be a blazar, but could be a stellar variable of some kind. (The other radio sources are all clearly AGN, and are likely to be blazars.)

\begin{figure*}
\centering
\includegraphics[width=0.95\textwidth,angle=0]{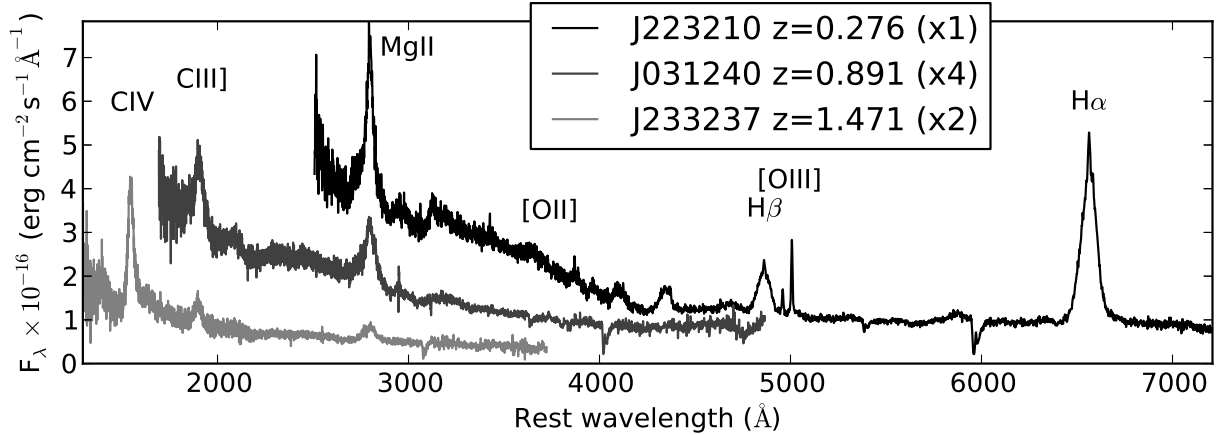}

\caption{\it\small Three example AGN spectra, taken from the lower, middle, and upper redshift ranges of our sample. The upper spectrum shows clear Balmer lines all the way through to $H\epsilon$ and possibly even $H\zeta$, but only $H\alpha$ and $H\beta$ are labelled. Note that the spectra are plotted against the AGN rest wavelength, rather than the observed wavelength, but the fluxes are observed fluxes, i.e. per unit observed wavelength.}
\label{fig:trispectrum}
\end{figure*}

Overall, we have a sample of 39 extremely variable AGN (including J025633) with spectroscopic information. With the exception of J025633, they are all broad-line AGN. The median redshift is $z=0.7$, and they cover the range $z=0.28$ to $z=1.99$. Three examples, at low, middling and high redshift, are shown in Fig. \ref{fig:trispectrum}. We see $MgII$ in almost all objects, $CIII$ and $CIV$ in higher redshift examples, and very clear Balmer series in lower redshift examples. At first glance, they look like fairly normal quasars, but to quantify this we have measured fluxes for some key lines. 

Table A5 tabulates some measurements of emission line strengths for 37 objects (not including J025633, which is featureless, and J105040, which has only a low-state SDSS spectrum). The fluxes were measured by fitting a polynomial continuum to line-free regions, and subtracting this fit, and integrating the remaining flux. Because of the redshift range, we see different combinations of lines; however broad Mg II~$\lambda 2798$ is seen in all objects, so we take this as a representative flux for the Broad Line Region (BLR). To represent Narrow Line Region (NLR) strength we used both [OII]~$\lambda 3727$ and [OIII]~$\lambda 5007$. [OII] is in the visible range for more objects, but is often undetected in our spectra, and is more likely to have a significant contribution from star forming activity in the host galaxy, whereas [OIII] will almost always be dominated by the AGN. In all cases, as well as line fluxes, we tabulate rest-frame equivalent widths as these will be much less susceptible to any flux calibration issues. We examine the spectroscopic properties of our sample a little more closely in section \ref{sec:lines}.


\section{Analysis}\label{sec:analysis}

\subsection{AGN and host luminosities}\label{sec:lums}

We calculated absolute magnitudes for the AGN transients, using the measured spectroscopic redshifts and a standard concordance cosmology  with $H_0=72\;$ km s$^{-1}$ Mpc$^{-1}$ and $\Omega_{tot}=1$. We used the peak $g$-magnitude, and calculated k-corrections assuming $F_\nu \propto \nu^{-\alpha}$ with $\alpha=0.5$. The AGN transients then have median $M_r=-23.87$. This is typical of a low-luminosity quasar. However, given the typical outburst amplitude of 2 magnitudes or more, the pre-outburst absolute magnitude is more like that of a Seyfert galaxy. 

For the AGN hosts, we assumed colours typical of the red sequence: $u-g=1.8$, $g-r=0.9$, $r-i=0.42$, $i-z=0.35$. The pre-outburst SDSS colours are sometimes bluer than this, but this may represent a mixture of red host and weak AGN, with the AGN being more significant in $g$ and $u$. Using these colours, the observed AGN spectroscopic redshift, where known, and the SDSS $r$-band magnitudes, we find absolute magnitudes with a median value of $M_r=-22.97$. This is an extremely large galaxy, even by AGN host standards (see e.g. \citet{Heckman2014}.) The observed Petrosian radii of our targets in the SDSS data is typically $\sim 1.5^{\prime\prime}$. At the median redshift $z=0.7$ this corresponds to a radius of 10.5 kpc, consistent with being a large galaxy. Alternatively, the $r$ magnitude may have a large AGN contribution; or the SDSS object may have a contribution from a lower redshift foreground galaxy, as we discuss below.

All except two of our targets were morphologically classified as galaxies in SDSS DR7 and have photometric redshifts, based on standard template fitting. (The photo-zs are listed in Table A4). For the objects where we have spectroscopic redshifts, the photometric and spectroscopic redshifts agree well for the known SNe, but are almost always strongly discrepant for the AGN, in the sense that the photometric redshift is always smaller. The typical photometric redshift is $z\sim 0.25$, and the template fitting requires a late type (Sc) galaxy in most cases. The most likely reason for the redshift discrepancy is that at the SDSS epoch our targets were actually a mixture of galaxy and AGN colours, with the AGN weaker than the galaxy at $r,i,z$ but similar strength at $u,g$. 

At the faint magnitudes of our targets, the SDSS morphological classification as galaxies may be unreliable. There is some evidence in the SDSS database that this is the case; although the {\em probPSF} parameter (which measures the probability of being consistent with the PSF) is always 0, the {\em probPSF} value for individual bands is sometimes set to 1. Another test is to compare the {\em cmodel} magnitude with the {\em PSF} magnitude. A little experimentation with stars of a similar magnitude in the same fields shows that the difference between {\em cmodel} and {\em PSF} magnitudes is well centred on zero, and usually within 0.05 magnitudes. For our objects, the {\em cmodel} magnitude is always brighter than the {\em PSF} magnitude, but sometimes only by $\sim 0.2$ magnitudes, and occasionally even less.  Overall then, it looks like our objects are resolved in the SDSS epoch imaging, but probably marginally so, and quite possibly resolved in some bands and not others. Finally, we note that in DR9, the morphological classification for 8 out of our 76 objects had changed to being starlike (see Table A4), directly confirming that many of the classifications are marginal.

A second possible reason for the discrepancy between photometric and spectroscopic redshifts in at least some cases is that that the photometric redshift is correct, and we are seeing an intervening foreground galaxy. Such foreground objects will be galaxies drawn randomly from the galaxy luminosity function, rather than a flux-weighted sample, and so will typically be fairly small and blue late-type galaxies. We calculated absolute magnitudes using the observed $r$-magnitude and k-corrections based on typical blue cloud colours: $u-g=1.2,\; g-r=0.04,\; r-i=0.3,\; i-z=0.05$. (Colours are referenced to $z=0.1$, but observed at a variety of redshifts.) Using the individual SDSS photo-z values we then find a median absolute magnitude $M_r=-19.72$, consistent with being a late type spiral half a magnitude or so below $L_*$ . This is at least roughly consistent with what one would expect for a randomly selected line of sight, as opposed to the $L_*$ galaxies that dominate flux limited surveys.

\subsection{AGN line strengths}\label{sec:lines}


\begin{figure}
\centering
\includegraphics[width=0.45\textwidth,angle=0]{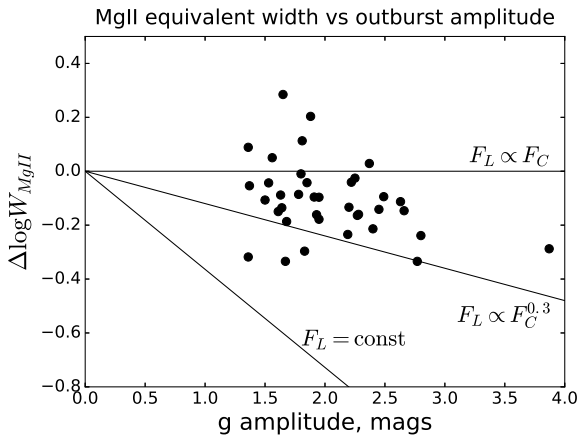}

\caption{\it\small Deficit in MgII equivalent width (EW), compared to the prediction from the relation between EW and luminosity for normal AGN given by \citet{Dietrich2002}. MgII seems to be weaker than normal in our hypervariables, with the deficit loosely correlated with the amplitude of the outburst.}
\label{fig:MgIIEW}
\end{figure}


To a first approximation, the spectra of our objects look like quite normal AGN - blue continuum, strong broad emission lines plus some narrow lines. However, they seem to have somewhat weak lines, as we discuss below. The relative strength of line to continuum could be an important diagnostic for the physical cause of extreme variability - for example how the broad lines respond to continuum variations must give us clues about the structure and formation of the BLR (e.g. \citep{Elitzur2014,Peterson2006,Korista2004,Goad1993}). Narrow lines on the other hand should remain constant, and so could tell us whether the low state or the high state is the normal one.

We first examine MgII, as this the line we see most consistently. The MgII line in AGN has a clear ``Baldwin effect'', i.e. the equivalent width depends on luminosity \citep{Dietrich2002}. We start by predicting the expected equivalent width $W_{pred}$, based on the Dietrich relation and the current observed luminosity (i.e. the high-state luminosity), calculated at a fixed wavelength of 1450\AA\ using an SED with $\alpha =-0.5$.  We then compare this to our observed rest-frame equivalent width $W_{obs}$ and calculate the deficit as $\Delta\log W = \log W_{obs} - \log W_{pred}$. The majority of objects (30/36) have $\Delta\log W <0$, i.e. have a deficit compared to expectation (Fig. \ref{fig:MgIIEW}.) There is considerable scatter. The median is $\log W=-0.11$ corresponding to a 30\% effect, but there are deficits up to a factor of 2. What if we had used the low-state luminosity to place our objects on the Dietrich et al relation? Generally, we do not know this, but for example, if an object is now a factor 10 brighter than its normal value, the correct value of $\log W_{pred}$ would be smaller by 0.1. The typical deficit may then be more like $\Delta\log W \sim -0.2$ i.e. a 60\% effect.

Regardless of the scatter in the EW relation, we have 6 points above the line and 30 below, which on simple binomial probability very strongly rules out our objects having normal equivalent widths for their observed luminosities. There is some weak evidence for a dependence on the $g$-band amplitude of the outburst (see Fig.\ref{fig:MgIIEW}), in the sense that larger amplitude events have weaker lines. For $\Delta g<2.0$, 5/21 objects are actually above the line; for $\Delta g>2.0$, only 1/12 objects is above the line. 

What would we expect to see during the outburst? This depends on how the line responds to the continuum change. The MgII line is known to have a very low responsivity (e.g. \citet{Goad1993}). A recent observation by \citet{Cackett2015} shows a factor two UV continuum change over several months while the MgII line changes by at most 20\%. If the MgII flux stayed constant during our continuum outburst, we would predict objects following the lower track in Fig. \ref{fig:MgIIEW}, which is strongly ruled out. So what we are seeing does not look like normal ``reverberation'' style changes, for which we would have expected very distinct equivalent width changes, and so very weak MgII in the high state.

However, we are studying much longer timescales than usual in reverberation studies. Timescales of many years, as opposed to weeks-months, may well be comparable to the dynamical timescale in the broad-line region, so the BLR may physically respond, as opposed to simply having a changing illumination. A recent study of several ``changing look'' AGN by \citet{MacLeod2016} shows the line flux in at least one well observed case being {\em proportional} to continuum flux over large long term changes. Such proportional changes would predict the upper track in Fig. \ref{fig:MgIIEW}. This is also ruled out as the typical case. The middle track in Fig.\ref{fig:MgIIEW} shows the expectation if the high-state to low-state ratio of line fluxes, $R_L$, follows the continuum ratio $R_C$ as $R_L=R_C^\gamma$, with $\gamma=0.3$ chosen simply to roughly bracket the points. The data points are spread between the upper two lines. It seems that the line flux responds to the continuum, but with smaller amplitude - for example, for a factor ten continuum change, the line flux has varied by somewhere between a factor two and a factor ten.

\begin{figure}
\centering
\includegraphics[width=0.45\textwidth,angle=0]{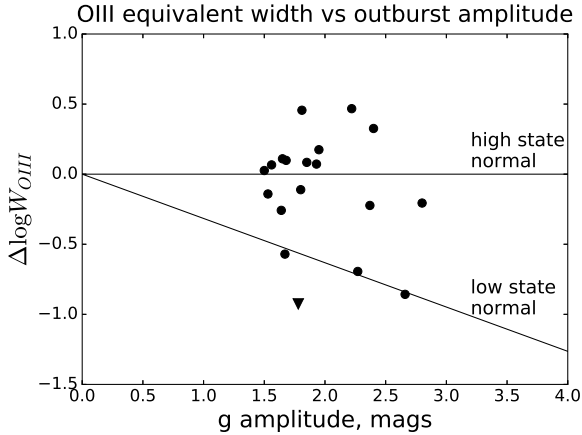}

\caption{\it\small Deficit in OIII equivalent width (EW), compared to the prediction from the relation between EW and luminosity for normal AGN given by \citet{Zhang2013}. Note that this test is only possible for the lower redshift/luminosity part of our sample.}
\label{fig:OIIIEW}
\end{figure}

In a similar way, in Fig \ref{fig:OIIIEW} we show the deficit in [OIII] equivalent width, $\Delta\log W_{OIII}$, using the prediction from the relation between EW and luminosity found by \citet{Zhang2013}. We use [OIII] rather than [OII] because most of our [OII] measurements are upper limits, but not at a low enough level to be strongly constraining, and because [OIII] is a better indicator of AGN activity, as opposed to star formation. It should be noted however that we only measure [OIII] in the lower redshift range, $z<0.8$, and hence lower luminosity, part of our sample. In Fig \ref{fig:OIIIEW} the upper line shows the prediction on the assumption that the current high-state is the normal one, with the SDSS data point being anomalously low, for example because of an extinction-event at that epoch. The lower line shows the prediction if the SDSS epoch low-state is the normal one, with the current state representing an outburst. Intriguingly, the points seem to be spread between these two lines. One interpretation could be that our objects are continually fluctuating at this extreme level, so that [OIII] strength reflects a time average between the upper and lower states.


\subsection{Ultra blue objects}\label{sec:ultrablue} 

In Section 3.1 we noted that while the majority of our targets have normal quasar colours, a significant minority (14\%) are ultra-blue, with $u-g<-0.05$. With the spectroscopic results in hand, we can see the probable cause. Of our 11 ultra-blue objects, 6 have spectra. They are at a variety of redshifts, but in five out of six cases there is a strong emission line (CIV, CIII] or MgII) close to the centre of the $u$-band, easily enough to distort the $u-g$ colour. 
It is likely then that this is also the cause of the ultra-blueness for the five objects without spectra.

Emission-line effects on broad-band colours are of course well known (see e.g. \citet{Schmidt2012}), but in section 3.1 we stressed that the fraction of ultra-blue objects was significantly larger than in SDSS. Almost certainly this is because of the difference in redshift distribution.

\subsection{Colour changes}\label{sec:colour}


\begin{figure}
\centering
\includegraphics[width=0.5\textwidth,angle=0]{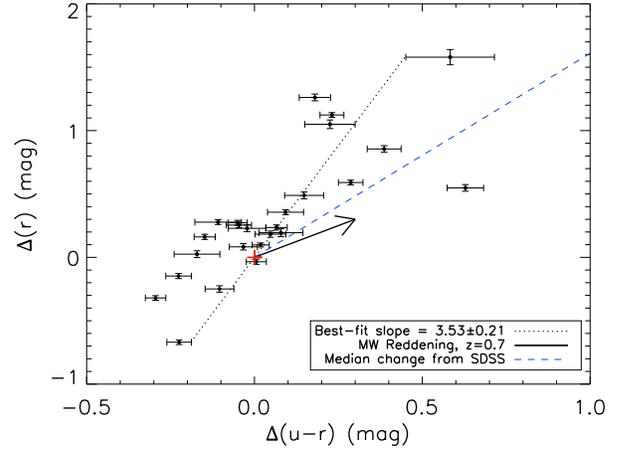}

\caption{\it\small Net colour changes during our period of monitoring, for objects with a consistent trend. Negative $\Delta r$ refers to rising objects. The dotted line shows a simple least squares trend line fitted to the data, minimising the deviations in both $r$ and $u-r$. The arrow shows a standard reddening vector of length corresponding to $A_v=1$. The dashed line indicates the change in the sample median values of $r$ and $u-r$ between the SDSS and PS1 epochs. If the PS1 peak epoch represents pure transient, and the SDSS epoch pure host galaxy, then the net colour should move along this line during the outburst.}
\label{fig:ur-r}
\end{figure}

\begin{figure}
\centering
\includegraphics[width=0.5\textwidth,angle=0]{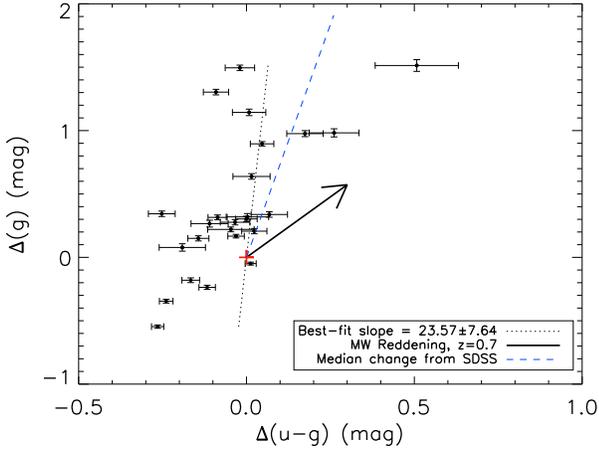}

\caption{\it\small Net colour changes during our period of monitoring. As for fig \ref{fig:ur-r}, but now for $u-g$ versus $\Delta g$. The trend line is consistent with no change in colour. Note that the best fit may be not exactly vertical partly as a consequence of forcing the line to pass through the 0,0 point. }
\label{fig:ug-g}
\end{figure}


An important test of possible physical models will be whether there are colour changes during the extreme variability we have seen - for example, changing extinction should produce a strong colour change. First, we look at the difference between the SDSS epoch and the PS1 transient trigger, where there are  clear colour changes. Taking all the objects classified as definite or likely AGN (54 targets), we find median values of $\Delta(g-r)=0.41, \Delta(u-r)=1.05, \Delta(u-g)=0.26$, for the median changes of $\Delta g=1.91$, $\Delta r=1.66$, and $\Delta u=2.41$. These large colour changes are very likely because the SDSS epoch was galaxy-dominated, whereas the PS1 epoch was AGN-dominated. As we saw in section 3.1, the colours of the SDSS objects are mostly consistent with galaxy colours, but sometimes a little on the blue side, suggesting either late-type hosts or the presence of a weak AGN component, which of course would be stronger in the blue than in the red.

In Figs \ref{fig:ur-r} and \ref{fig:ug-g} we look at colour gradients during our monitoring with the Liverpool Telescope. To simplify this, we selected objects that have been clearly consistently rising or falling during our monitoring period, restricting us to 25 targets. 
The $u-r$ colour gives the most leverage, and Fig. \ref{fig:ur-r} shows a clear colour trend. However, this seems to be consistent with the long term colour trend due to changing AGN-galaxy mixture. Because the host galaxy will be red and the transient blue, as the transient fades, the $r$-band light will become dominated by the galaxy sooner than the $u$ or $g$-band light.
The $u-g$ colour should therefore be a better test of the behaviour of the transient. Fig \ref{fig:ug-g} shows a considerable scatter, but the best fit trend line is close to vertical, i.e. our light curves are consistent {\em on average} with achromatic changes. Note that for both $u-g$ and $u-r$, the colour changes or lack thereof seem inconsistent with simple reddening changes. 

\subsection{Variability in context: extreme variables in Stripe 82}
\label{sec:context}

Our selection method required objects to be morphologically classified as galaxies in the SDSS epoch. Do objects already classified as AGN occasionally show similar extreme variability? All AGN, including luminous quasars, vary, but this variability is typically only a few tenths of a magnitude, and is wavelength dependent \citep{Giveon1999, Hawkins2003, Hawkins2007, VandenBerk2004, DeVries2005, Sesar2006, MacLeod2010, Schmidt2012}. Examples of extreme variability are known, e.g. \citep{Lawrence1977,Penston1984,Shappee2014,LaMassa2015,MacLeod2016}, but how common is this? 

The best information to date comes from the study of \citet{MacLeod2012}, who looked at repeat SDSS observations of the  SDSS quasars. McLeod \et\ fit population models to the histogram of $\Delta m$ and predict numbers of variables to various survey depths. For quasars to PS1 depths, this analysis predicts one quasar in $\sim 10^5$ to have $|\Delta g|>1.5$.  We examined this issue more directly by searching their online dataset for rare extreme variables. The MacLeod catalogue contains 33,881 quasars with at least two epochs in at least one band. Note that where more than two epochs were available, most notably for Stripe 82 objects, the longest time difference was used by MacLeod et al. From this sample, we removed objects with g-band error larger than 0.15 mag, and those with only one g-band observation, leaving 33,418 objects. For these, the mean absolute value of $|\Delta g|$ is 0.14 mag, with a standard deviation of 0.17 mag. Objects with large variability do exist, but they are very rare. There are 130, 19, and 6 objects respectively with $|\Delta g|>1.0, 1.5, 2.0$, and only 1 with $|\Delta g|>2.5$. 

We note that the majority of these extreme variables come from the objects in the Stripe 82 survey. Out of our cleaned sample of 33,418 objects, 9078 are from Stripe 82; however 15 out of 18 objects with $|\Delta g|>1.5$ are from Stripe 82. Very likely this is because the typical time difference between observations is longer for Stripe 82. For the non-Stripe 82 objects in the MacLeod et al sample, the mean difference between the two observations used is 1.3 years, whereas for the Stripe 82 objects it is 8.6 years. The variability of AGN is known to increase with longer timescales, representing something like a damped random walk, although with some debate about whether we have or have not reached the knee of the structure function \citep{DeVries2005, MacLeod2012, Morganson2014}. 

\begin{figure}
\centering
\includegraphics[width=0.30\textwidth,angle=-90]{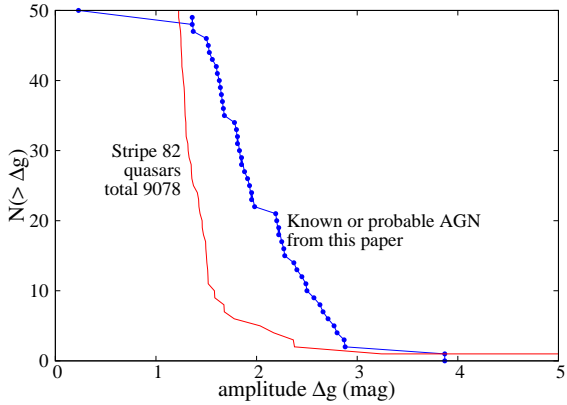}

\caption{\it\small Comparison of variability amplitude histograms between Stripe 82 and the sample of this paper. For Stripe 82, the $\Delta g$ is the change between first and last epochs, as in the catalogue of \citet{MacLeod2012}, as explained in the text.}
\label{fig:hist-s82}
\end{figure}

Overall the Stripe82 dataset seems a much better comparison to our SDSS-vs-PS1 sample. Fig. \ref{fig:hist-s82} compares the cumulative histogram of amplitudes seen in the Stripe 82 sample to the amplitudes seen in our PS1-FGSS sample. We are clearly seeing a very rare tail of variability in AGN, and it could be even rarer than is apparent from this comparison - the PS1-FGSS amplitudes are lower limits, because the SDSS epoch was likely galaxy dominated. All in all, perhaps somewhere in the range 1 in 1,000 to 10,000 AGN show the kind of extreme variability over a decade that we have been studying.

We also see several times more such ``hypervariable'' objects than are in the Stripe 82 sample. This is because we are drawing from a larger potential pool of AGN, have a somewhat longer baseline, and are sensitive to objects which were extended at the earlier epoch. Our starting sample was SDSS galaxies to a depth of $g\sim 22$. From the tables in \citet{Yasuda2001} we estimate the density of SDSS galaxies to a depth of $g=22$ to be 5000 sq.deg$^{-1}$. How many of these galaxies host an AGN? As we discuss in section 4.1, the quiescent luminosities of our objects are similar to those of classic Seyferts (rather than for example dwarf Seyferts) and so perhaps 1\% of galaxies host such an AGN. Over the SDSS footprint (11,667 sq.deg.) and a ten year time gap, we expect to see somewhere in the range 50-500 transients. This calculation is obviously very crude, but what we have seen is indeed roughly consistent with the PS1 hypervariables being the same population that we see in the extreme tail of the Stripe 82 quasars.

\begin{figure}
\centering
\includegraphics[width=0.30\textwidth,angle=-90]{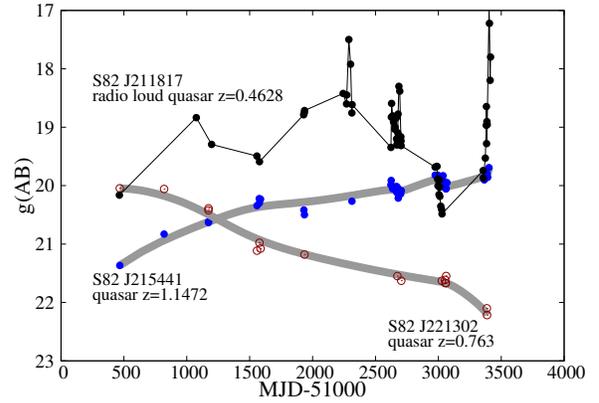}

\caption{\it\small Ten year light curves for selected Stripe 82 hypervariables. The grey curves are smooth polynomial Bezier curves meant only to guide the eye. The thin solid line for J211817 is simply a ``join the dots'' line, meant to illustrate the erratic variability of this object compared to the other two. }
\label{fig:lc-s82}
\end{figure}

Do the Stripe 82 hypervariables behave the same way as the PS1 hypervariables? Table \ref{tab:S82} lists the 15 Stripe 82 hypervariables with $|\Delta g| > 1.5$, and Fig. \ref{fig:lc-s82} shows example light curves. First we note that of the three objects that are radio detected, two have extremely erratic light curves, and so are very likely blazars. Of the remainder, almost all are smoothly changing over a decade, with two objects showing some kind of second peak. This is very similar to our PS1 hypervariables, except that in the Stripe 82 sample we are more sensitive to downward as well as upward changes. In fact, we see more objects going down than up (9 vs 4). However, on the assumption of symmetrical light curves, objects in their high state are more likely to be in the SDSS quasar sample than objects in their low state, so a bias towards declining light curves is expected. Overall it seems very likely that we are seeing the same phenomenon in a handful of hypervariable Stripe 82 quasars that we have seen in our larger sample of PS1-vs-SDSS transients.

\begin{table*}
\caption{Hypervariables in SDSS, Stripe 82 selected as explained in the text. The magnitudes are PSF magnitudes. The ``radio'' flag indicates whether the object has an entry in the NVSS catalogue.}
\begin{tabular}{|l|l|l|l|l||l|l|l|}
\hline
SDSS ID & $\Delta g$ & $z$ & radio & $\Delta{\rm MJD}$  & 
$g_1$& $g_2$ & light curve \\
\hline
  J001016.22+004713.3 & 1.58 & 1.181 & 0 & 2215.9 & 19.77 & 21.36 & down, 2ndpeak\\
  J001130.40+005751.7 & 3.24 & 1.4915 & 1 & 3321.0 & 17.88 & 21.12 & erratic\\
  J001420.44-003620.3 & 2.38 & 0.9589 & 0 & 3330.9 & 19.83 & 22.21 & smooth down\\
  J013815.05+002914.0 & -1.58 & 0.9402 & 0 & 3310.0 & 21.22 & 19.64 & smooth up\\
  J032544.82-011028.5 & -1.51 & 1.2727 & 0 & 3321.0 & 21.18 & 19.67 & smooth up, flat top\\
  J032946.99+000002.7 & 1.78 & 1.4147 & 0 & 3333.9 & 20.22 & 22.00 & smooth down\\
  J033931.17+002905.6 & 1.51 & 0.9859 & 0 & 3310.0 & 20.44 & 21.96 & smooth down\\
  J034137.03-000915.5 & 2.03 & 0.602 & 0 & 3313.0 & 20.28 & 22.32 & smooth down\\
  J205518.60-005635.0 & 1.52 & 0.9237 & 0 & 2218.0 & 20.36 & 21.88 & down, 2nd peak\\
  J211817.39+001316.7 & -2.36 & 0.4628 & 1 & 2951.0 & 20.16 & 17.80 & erratic\\
  J215441.95+001008.0 & -1.68 & 1.1472 & 1 & 2934.9 & 21.37 & 19.69 & smooth up\\
  J221302.57+003015.9 & 2.17 & 0.763 & 0 & 2919.9 & 20.05 & 22.22 & smooth down\\
  J233317.38-002303.4 & -1.50 & 0.5129 & 0 & 1880.3 & 21.22 & 19.72 & sharp rise\\
  J234855.04+002539.1 & 1.52 & 1.2745 & 0 & 2937.9 & 19.42 & 20.93 & smooth down\\
  J235439.14+005751.9 & 1.68 & 0.3896 & 0 & 3321.0 & 19.05 & 20.73 & smooth down\\
\hline
\end{tabular}
\label{tab:S82}
\end{table*}


\subsection{Notes on individual objects}\label{sec:individuals}

Before proceeding to discuss the nature of the slow-blue hypervariables, we present some short explanatory notes on a handful of individual objects.
 
{\bf J012514}. This is morphologically classed as a galaxy in both SDSS DR7 and DR9, but our spectrum (in quiescence after a fast decay) shows weak H$\alpha$ at $z=0$, and it therefore is presumably some kind of stellar variable. The $u$-band amplitude was 4.7 magnitudes, which would be large but not unprecedented for a CV. It is not unnusual for a flare star \citep{Kowalski2009}, but such an extended duration is not normal. 
 
{\bf J025633}. This is one of the brightest transients, which varied erratically during the outburst. 
At peak it was very blue and had a featureless spectrum. Unforunately we therefore failed to get a redshift. It is a moderately strong radio source. As well as NVSS, it is detected in the GB6 4.8GHZ survey \citep{Gregory1996}, and in the CLASS survey at 8.4GHz \citep{Myers2003}. It has a flat radio spectrum and is therefore very likely a blazar. We note that in SDSS DR9 the morphological classification was changed to stellar, and the quiescent colour is very red, $u-g=1.01$. 

{\bf J044918}. Like J012514, this is morphologically classed as a galaxy in both SDSS DR7 and DR9, but our spectrum  shows weak H$\alpha$ at $z=0$, and it therefore is presumably some kind of stellar variable.  

{\bf J061829}. This was one of two objects (along with J083544) classified as stellar in SDSS DR7, but which satisfied our other criteria, which we selected as a comparison. Our WHT spectrum shows it to be a cataclysmic variable, with strong Balmer line emission and a flat Balmer decrement. It has a very large outburst amplitude (5 magnitudes) and erratic variability on top of a slow decline.

{\bf J081916}. This is one of two objects (along with J105040) in our sample, which despite being morphologically classified as a galaxy in SDSS DR7, has an SDSS era spectrum, taken because the object was a ROSAT target. In Feb 2002, the SDSS spectrum shows a galaxy dominated continuum, strong [OIII], narrow H$\beta$, and weak broad MgII, and so could be classified as Seyfert 1.9. Our 2013 spectrum shows a much stronger and bluer continuum, broad H$\beta$, and strong broad H$\alpha$, a normal looking Type 1 AGN. This object is then an example of a ``changing look'' AGN (c.f. \citet{Shappee2014, LaMassa2015, MacLeod2016}).

{\bf J083544}. This is the second of two objects (along with J061829) classified as stellar in SDSS DR7, but which satisfied our other criteria, which we selected as a comparison. Our WHT spectrum shows it to be a quasar at $z=1.327$. It is 2 magnitudes brighter than the SDSS epoch, and still rising.

{\bf J094309}. The radio source associated with this AGN is highly variable - the VLSS 20cm flux is two orders of magnitude larger than the NVSS 20cm flux. It is therefore very likely a blazar.

{\bf J105040}. This is the second of two objects (along with J081916) in our sample, which despite being morphologically classified as a galaxy in SDSS DR7, has an SDSS era spectrum, taken because the object was a ROSAT target. In Jan 2004 this was a low redshift AGN with strong [OIII], broad H$\beta$ with a strong narrow core, and a continuum with significant galaxy contribution. Unfortunately we do not yet have a PS1 epoch spectrum. Although not detected as a CRTS transient, it is clearly variable in the CRTS data.

{\bf J133155}. This object remains of unknown type; unfortunately we do not have a spectrum.  It was one of the bluest transients in our sample, with $u-g=-0.54$, and had one of the largest outburst amplitudes, with $\Delta g=2.71$ and $\Delta u=3.64$. The decay rate (0.58 magnitudes/month) was much faster than known AGN transients, with the exception of the (also very blue) TDE candidate PS1-10JH \citep{Gezari2012}. It is not as fast or erratic as the known emission line stars in our sample. The fast decay is consistent with being a supernova, but it could also have been a tidal disruption event. 

{\bf J160329}. The radio source associated with this AGN has a flat spectrum between FIRST and NVSS at 20cm, and GB6 at 6cm. It is therefore very likely a blazar.

{\bf J202823}. Like J012514 and J044918, this object was morphologically classed as a galaxy in both SDSS DR7 and DR9, but our spectrum shows it to have $z=0$ emission lines, so that it is presumably a stellar variable of some kind. The spectrum is much richer than those of J012514 and J044918. It shows strong Balmer lines with a flat decrement, as well as strong He lines, both HeI$\lambda 5876$ and HeII$\lambda 4686$. It is probably therefore a cataclysmic variable of some kind.

\section{Discussion: what are the slow blue nuclear transients?}\label{sec:nature}

We have found a class of luminous AGN which have brightened by an order of magnitude since a decade ago, and are now found to be steadily changing, mostly fading and sometimes increasing in flux. (From here on we ignore the handful of erratically varying radio-loud objects.) Archival light curves for some show that they have smoothly evolved over ten years. Spectroscopically, these objects seem to be normal AGN, except that the MgII broad lines seem to be somewhat weaker than expected for their luminosity. Because we selected objects classified as galaxies in the past, it is unknown how bright the AGN was a decade ago, or indeed whether it was there at all. However we have identified an analogous set of smoothly evolving hypervariable quasars in the SDSS Stripe 82 dataset, that have clearly been AGN throughout. We examine four possibilities that may explain the observed behaviour: (i) tidal disruption events, (ii) extinction events; (iii) eruptive accretion flares, and (iv) foreground microlensing.

\subsection{Tidal disruption events}\label{sec:TDE}

Our original aim was to locate examples of TDEs - dormant black holes which come temporarily to life when a star passing close to the black hole is torn apart by tidal forces. Good candidates have been found for such events, both in the optical/UV \citep{Gezari2008, Gezari2009, Gezari2012, VanVelzen2011, Chornock2014} and in X-rays \citep{Brandt1995, Grupe1995, Bade1996, Halpern2004, Komossa2004, Komossa2012}. Could our slow blue transients be examples of TDEs?

Initial models suggested that such events will be extremely blue, have peak luminosities of the order $10^{37}$ W, and after a fast rise will decay on a timescale of months with flux following $t^{-5/3}$ \citep{Rees1988, Evans1989}. Recent work shows that the details may be rather more subtle \citep{Lodato2011, Guillochon2013}, but the broad characteristics are fairly clear. Such events will occur much more frequently around relatively low mass black holes ($\sim 10^{6-7} M_\odot$) because of the steeper force gradient, giving an Eddington limited luminosity of $L_{bol}\sim 10^{37-38}$ W. Most disrupted stars will be of significantly less than a solar mass, and only a fraction of the mass will be accreted. Using $m_*=0.3M_\odot$, accretion fraction $f=0.5$, and accretion efficiency $\mu=0.1$ gives a total flare energy of $E = 3\times 10^{45}$J 
and at the Eddington luminosity, the timescale to consume half the the material is 
$t_{1/2} = 122$ days. 

These characteristics agree fairly well with the events seen by \citet{Gezari2009, Gezari2012}, and \citet{VanVelzen2011}. However, they differ substantially from the events we have discussed in this paper. Using standard bolometric corrections from e.g. \citet{Elvis1994} or \citet{Richards2006}, our objects have peak luminosities of $L_{bol}\sim 10^{39}$ W. At the Eddington limit, this requires black holes of mass $M>10^8 M_\odot$. TDEs should occur rather rarely in such massive black holes, because for most star types, the tidal radius is inside the event horizon. The typical two folding timescale for our hypervariables is $t_{1/2}\sim 900$ days, significantly longer than expected for TDEs. More strikingly, if we combine the high luminosity with the long timescale, we find that the flare energy is typically $E\sim 10^{47}$J. At accretion efficiency $\mu=0.1$, if the accreted fraction is $f\sim 0.5$, then the star consumed must have had $m_*\sim 10 M_\odot$.  

Similar values were found by \citet{Meusinger2010} to explain the event in the quasar seen behind M31 - a large star tidally disrupted by a large black hole. While this may be a plausible explanation for a specific event, it seems unlikely to be the explanation for the majority of the Pan-STARRS nuclear transients.

\subsection{Extinction events}\label{sec:extinction}

Another possibility is that the large change in flux we have seen is due to a change in extinction, with the PS1 high state being the normal state, and SDSS having seen the object during a period of high extinction. Extinction variations have occasionally been discussed as the possible cause of rare large amplitude changes in relatively nearby AGN, (e.g. \citet{Tohline1976, Goodrich1995, Aretxaga1999, LaMassa2015}). It is commonly believed that the parsec scale obscurer in AGN is patchy or clumpy \citep{Krolik1988, Nenkova2002}; motion of these clumps across the line could be the simplest cause of such changes. The simplest picture is where the size of the obscuring cloud is larger than the optical source size, producing a broad flat bottomed eclipse, with the source flux recovering as the trailing edge of the cloud moves the face of the source. If we assume that the obscuring cloud is in Keplerian motion at a radial distance $R_{cl}$ from the central black hole of mass $M$, and the optical source diameter is $D_{opt}$, then the recovery timescale would be

$$ t = 51.0 {\rm\; days\;} 
\left( \frac{M}{10^8 M_\odot} \right)
\left(  \frac{D_{opt}}{100 R_S} \right) 
\left( \frac{R_{cl}}{1000 R_S} \right)^{1/2}
$$

where $R_S$ is the Schwarzschild radius. For our spectroscopic AGN sub-sample, the median luminosity at 1450\AA\ is $\lambda L_\lambda \sim 10^{38.0}$ W. Note that in this model, the peak luminosity would be the normal one. For a bolometric correction of 5 \citep{Richards2006} and an Eddington fraction of $f=0.1$, this implies a black hole mass of $4\times 10^8 M_\odot$. A timescale of 10 years therefore suggests a cloud at $3.2\times 10^5 R_S \sim 12.5$ pc. This is much too large to be a dust-bearing BLR cloud, but is a very plausible distance for where the bulk of the geometrically thick obscuring material resides. For example, based on equation 1 of \citet{Lawrence2010}, the SED peak at $\sim 10\mu$m for an object of this luminosity would come from $\sim 1.4\times 10^5 R_S$.

The timescale is therefore quite plausible. However, the lack of colour changes in the period of our monitoring (see Section \ref{sec:colour}) argues strongly against a simple change of optical depth. The smoothness of the changes we see suggests that if we are looking at an extinction event, what we see must be more like a covering factor change, as an opaque cloud moves across the face of the source (i.e. an ``unveiling'' event).  A more realistic model would be between these two extremes, with both covering factor and optical depth changes; it would seem surprising not to see erratic changes. Also, an opaque eclipse model would normally lead to a flat-topped light curve; to come to a peak and then decline again slowly, the cloud size must be similar to the source size, and one would need two successive events. Overall, an extinction event seems not the best explanation of what we are seeing, but it can't be completely ruled out. 

\subsection{ Accretion events}\label{sec:accn}

It is clear that large amplitude changes are not normal behaviour for AGN (see section \ref{sec:context}), but not completely unprecedented (e.g. \citep{Khachikian1971, Tohline1976, Lawrence1977, Penston1984, Goodrich1995, Aretxaga1999, Bischoff1999, Shappee2014, LaMassa2015,MacLeod2016}, with perhaps one AGN in $10^4$ showing events as large as we have seen. It seems intrinsically unlikely that one AGN in every 10,000 varies in a different manner to all the others; however it is possible that every AGN has a short eruptive event of some kind once every 10,000 years. Spatial variations of X-ray emission from molecular clouds surrounding Sgr A* have been interpreted as a light echo tracing of past large amplitude temporal variations from the black hole in the centre of our own Galaxy, on timescales from a few years to thousands of years \citep{Ponti2013, Ryu2013, Clavel2013}.

It is hard to be confident in modelling such a possibility, as we donÕt yet understand normal AGN optical variability - it is fast, co-ordinated across wavelengths, and highly wavelength dependent in amplitude, all in contrast to expectation from simple accretion disc models (see \citet{Lawrence2012} and references therein). It is likely that the observed timescales - days to months - correspond to a thermal or dynamical timescale in the inner disc, with outer parts of the disc tracking variability in the inner parts through reprocessing of some kind
(e.g. \cite{Lawrence2012}). The viscous timescale is of the order $10^4$ years for a disc around a $10^8 M_\odot$ black hole (e.g.\citet{Frank2002}). This fits with the possible gap between events, but not the duration of the events we see.


Perhaps the event itself corresponds to crossing some critical point where the microphysics changes. For example, initially the disc may be cold, with a very low viscosity and accretion rate, but very slowly warming. When the temperature becomes warm enough for a modest ionisation fraction, the magneto-rotational instability (MRI) may quite suddenly switch on and greatly increase the viscosity and accretion rate. A model involving repeated changes of viscosity state due to the hydrogen ionisation instability was produced by \citet{Siemiginowska1997} and further developed by \cite{Hatziminaoglou2001}. From Figure 2 of the latter paper, it seems that to achieve a typical time between outbursts of $10^4$ years requires a black hole of mass $\sim 10^7 M_\odot$ , but the duration is then also $\sim 10^4$ years which does not fit our observed events. For any similar model, the switch on, and any subsequent evolution, is likely to be on timescales corresponding to the viscous timescale, i.e. of the order of tens of thousands of years.

One possibility is that the disc becomes empty in the inner regions, and is accumulating (cold) material beyond some truncation radius. The waiting time is due to the ionisation instability as discussed above, but following the transition the inner disc refills catastrophically on a dynamical timescale. For a black hole mass of $\sim 10^8 M_\odot$ this would require a truncation radius of $\sim 1000 R_S$. A related ``stall and refill'' model has been discussed by \cite{Grupe2015} in relation to the recurring extreme X-ray AGN transient IC~3599. 

Finally, it could be that what is needed is  for some kind of instability to generate energy but not to radiate this energy immediately - rather, it needs to be stored in the disc and then released all at once. Then the decay time would be the thermal (cooling) timescale of the disc. For standard disc models, this is of the order of the dynamical timescale divided by $\alpha$, the viscosity parameter, and could indeed be of the order of years \citep{Collier2001a, Kelly2009}. However one would expect that the colour would change greatly during the flare, which does not seem compatible with the lack of systematic evolution in $u-g$ discussed in section \ref{sec:colour}.

Overall, the idea of some kind of rare eruptive accretion flare that is very short compared to to the time between events, is plausible but hard to judge without more detailed models.

\subsection{Microlensing events}\label{sec:microlensing}

The fourth possibility for explaining large amplitude changes is microlensing by a star in a foreground galaxy. Microlensing has in the past been proposed as a general cause of quasar variability, due to intergalactic compact objects \citep{Hawkins1993,Hawkins1996,Hawkins2007}. We are not reviving that idea here, but considering the possibility of rare exceptional events.

Microlensing is well established as an explanation of differential variability between components of lensed quasars, where the macro-lensing is caused by the overall foreground galaxy potential, but each component suffers different amounts of micro-lensing as the light takes different paths through the galaxy \citep{Irwin1989,Eigenbrod2008,Morgan2010,Blackburne2011, Mosquera2011,Jimenez-Vicente2012,Jimenez-Vicente2014,MacLeod2015}. Here we are considering something something different - rather than ongoing low-level statistical variability caused by the overlapping magnification effect of many stars, we have rare temporary large amplitude events caused by passage close to a single star, or a caustic caused by a small number of stars. As we argue below, this will most often happen where the foreground galaxy is small, and so the macrolensing is modest, with the macrolensed components having a small angular separation. 

Microlensing events have been discussed as the explanation of large amplitude flares twice in the past. The first case is the blazar AO 0235+164 \citep{Stickel1988, Webb2000}, but this seems unlikely to be the correct explanation because the flaring repeated, and also made new radio structures \citep{Kayser1988}. The second case is that of a 3.5 magnitude 6 year flare in a quasar behind M31 \citep{Meusinger2010}. Meusinger at al consider two possibilities - that this was a microlensing event, or a TDE. They favour the latter, because of the rarity of such a high amplitude microlensing event, and because the light curve shape, while looking roughly like the classically expected cuspy shape (see eg \citet{Schmidt2010}), has a shoulder, requiring a two-star lens. On the other hand, the TDE explanation is also a little forced, requiring the disruption of a rather rare $10 M_\odot$ red giant in the presence of an already existing accretion disc.

What are the characteristics of an event that could explain what we see - large amplitude events with a timescale of several years? In the following paragraphs we use the following canonical numbers. First we assume an amplitude of a factor $A=10$, based on the median $\Delta u$. Next, we take the source to be $z_s=1$, at angular diameter distance 1652 Mpc, and we take the lens to be at $z_l=0.25$, roughly halfway in angular diameter distance. Finally, we take a representative lens mass of $m=1 M_\odot$, which is roughly the median of the mass-weighted stellar mass function of \citet{Chabrier2003}. (Lensing area is proportional to mass, so this is the mass for which half the lensing area is above/below).

{\bf Lens and source sizes}.
For a standard single lens light curve (see e.g. \citet{Schmidt2010}) the amplitude near peak is $A\sim 1/u_{min}$ where $u_{min}$ is the impact parameter scaled to the Einstein radius, i.e. $u_{min}=\Delta\theta_{\min}/\theta_E$. For $A=10$, $u_{min}=0.1$ and so 

$$\theta_{min} = 0.29 {\;\mu\rm as\;} \left(m/M_\odot \right)^{1/2}
\rightarrow R_{min}=7.19\times 10^{13} {\rm\;m\;}
$$

where $R_{min}$ is the corresponding physical size at the source plane. How does this compare to the likely source size? For a black hole of mass $M_H$ we have

$$R_{min}/R_S = 244 \left(M_H/10^8 M_\odot \right)^{-1} \left(m/M_\odot \right)^{1/2} $$

To a first approximation therefore the accretion disc will be unresolved and the broad line region (BLR) {\em partly} resolved, and so amplified, but by less than the continuum. This could be why we are seeing Mg II equivalent widths that are weaker than normal (see section 4.2) and gives us the fascinating prospect of measuring BLR structure (c.f. \citet{Sluse2012}). However, recalling that we are looking here at our canonical numbers, some events will resolve the continuum, and the peak of the light curve will hold information well below the scale of $R_{min}$. Furthermore, differential variability studies of strongly-lensed multiple quasars seem to indicate that continuum source sizes are several times bigger than simple accretion disc models (e.g. \citet{Morgan2010,MacLeod2015}), but this relies on statistical modelling of the stellar population in the parent galaxy. Direct model fitting of individual high amplitude events is therefore of considerable importance.

The lens size at the lens plane corresponds to 240 A.U. This means that we are almost always seeing the effect of single lenses, rather than the overlapping effect of many stars. (If say we are looking through a 5kpc column of stars with a space density of 1 pc$^{-3}$, a box size of 240 A.U. will on average have 0.007 stars in the line of sight). On the other hand, we should occasionally see double peaks - 50\% stars in the solar neighbourhood are in binaries, and while their separations cover a large range, the median value is of the order of a few tens of A.U. \citep{Duquennoy1991, Fischer1992}. We should also be sensitive to brown dwarf companions, and potentially even to extragalactic exoplanets, although these would produce very weak second peaks.

{\bf Timescales}.
For a characteristic timescale, we can use the crossing time of the lens across the source. The lensing star will be in motion within its galaxy with a typical value of perhaps 300 km s$^{-1}$, and the two galaxies will have peculiar velocities of a similar order, each in effectively random directions of course. For the net relative transverse velocity, we can use $v=300$ km$^{-1}$ as a characteristic value, and so find

$$ t_{ch} = 7.4 {\;\rm yrs\;} 
\left( v/300 \right)^{-1}
\left( u_{min}/0.1 \right)
\left(m/M_\odot \right)^{1/2}
$$

Note that this is the characteristic timescale near the peak, and is much shorter than the ``Einstein timescale'' often quoted. It is about right for our observed events. How often would such an event repeat? This depends on the surface density of stars through the foreground galaxy. We estimate a typical value using the analysis of \citet{Kauffmann2003} who suggest a characteristic stellar mass of $6\times 10^{10} M_\odot$ and half light radius $R_{50}\sim 3 $ kpc. We then estimate a rough typical surface density as $\Sigma = N/pi R_{eff}^2$ with $N=6\times 10^{10}$ and $R_{eff} = \sqrt{2} R_{50}$. Placing this at our canonical distance of $z_l=0.25$ this gives a characteristic angular radius of $1.1^{\prime\prime}$. If we think of a lens of the size above sweeping across a line of sight, we can find that the repeat timescale will be

$$t_{rpt} = 1354 {\;\rm yrs\;} 
\left( v/300 \right)^{-1}
\left( u_{min}/0.1 \right)^{-1}
$$

This would then give a duty cycle (fractional on-time) of $f_{on}=2 t_{ch}/t_{rpt} = 1.1\%$.

{\bf Number of flaring AGN}. 
Of course only some AGN will have a foreground galaxy in the line of sight. From the number counts in \cite{Yasuda2001}, there are roughly 5000 galaxies/sq.deg to $g=22$. If we take each of these to have radius $1.1^{\prime\prime}$, as in the above calculation, we estimate that the fraction of background AGN with a foreground galaxy is $f_{fg}=0.14\%$. Combining this with the duty cycle above, we find that the fraction of AGN we will see flaring at any time is $f_{fl}=7.8\times 10^{-6} \left(u_{min}/0.1 \right)^2$. 

How does this compare with general AGN variabilty?
If we extrapolate back to 1 magnitude changes, where we have reasonable statistics, roughly 1\% of Stripe 82 quasars vary by this much over a decade (see section 4.5), whereas we predict roughly 1 in $10^4$ AGN to be undergoing a microlensing flare. So it is unlikely that microlensing is the main cause of quasar variability at more modest levels, but it is possible that it dominates the extreme tail of variability. 

How many potential background AGN are there? Traditional optical quasar surveys are not helpful here because they do not go deep enough, and because for the Seyfert-like luminosities we are concerned with here the light is dominated by the host galaxy. The best information comes from deep X-ray surveys. The deepest such survey is the Chandra Deep Field South survey of \cite{Lehmer2012}. However, many of their sources are at lower luminosity and/or higher redshift than we are concerned with here. Keeping at $L_x\geq 2\times 10^{36}$ W and $z\leq 1.5$ we find about 2000 AGN/sq.deg. Putting this together with the SDSS DR7 area (11,667 sq.deg.) and the flaring fraction above, we predict that we should have seen 182 AGN currently in factor ten microlensing flares. Given the extreme roughness of all the estimates in this section, this is quite reasonable agreement with what we have in fact seen.

{\bf Light curve shape}.
Some of the ten-year light curves (J094511, J105502, J085759) show roughly the kind of peaky and symmmetric light curves one expects from a simple point-lens point-source system, especially given that intrinsic variability would be superimposed, and there may be flux-dependent offsets between the PS1/LT data and the CRTS data. J150210 seems too asymmetric. For the three-year light curves, we only have one side of the light curve. Some look consistent with the simple model, but some clearly don't, either because they are rather flat-topped (e.g. J031240) or because they have double peaks (eg J170845). However, although our canonical example has an unresolved accretion disc, some objects will be resolved, especially allowing for the evidence from multiple-quasar microlensing that accretion discs are $\sim 5$ times larger than predicted by standard models, e.g. \citet{Morgan2010}. Furthermore, although the host galaxy itself is likely producing only modest magnification, it will produce a shear that breaks the point-source singularity and makes a caustic-like magnification map leading to double peaks in a large fraction of cases \citep{Chang1984}. These issues are beyond the scope of the current paper, but quantitative modelling and testing of predicted light curves is a high priority for future work.

\subsection{Other possible microlensing examples}

Are there other possible examples in the existing literature of high amplification extragalactic micro-lensing events? The most obvious one is the flaring quasar behind M31 (Meusinger et al. 2010) as discussed in the previous section.

A second possible example is the TDE candidate PS1-10jh \citep{Gezari2012}. We note that the broad lines identified as HeII lines at $z = 0.17$ could also be identified with MgII and CIII] at $z = 0.97$. The shorter timescale (months) and the extremely large amplification (a factor of 200) are consistent with each other. However, overall the TDE explanation is probably still preferred. Firstly, the host galaxy absorption redshift seems secure, so on the microlensing hypothesis, the mis-identification of MgII and CIII] lines with HeII lines at the correct redshift would be a somewhat extraordinary coincidence. Secondly, for such a high amplification, the broad lines should be amplified far less than the continuum, predicting a very small equivalent width, which is not seen. Thirdly, the light curve seems asymmetric. This could be caused by a double star, as discussed above, but for a such high amplification, the double star would need to have a very small separation, which is once again unlikely.

Another interesting possibility is the superluminous supernova PS1-10afx \citep{Chornock2013}, at $z = 1.388$. This reached a peak luminosity of $4.1 \times 10^{37}$ W, but its colours did not look like other known superluminous SNe, and the spectrum looked like a normal SNIc, which should be 50 times less luminous. \citet{Quimby2013} suggested that the object is actually a SNIa lensed by a foreground object by a factor of thirty. A foreground object at $z=0.117$ was detected spectroscopically by \citet{Quimby2014}, who also derive a stellar mass of $\sim 10^{10} M_\odot$ from spectroscopic modelling. Magnification by a factor of thirty is very large for such a small galaxy, but \citet{Quimby2014} show it is statistically allowed, if the alignment is very good - roughly 0.02\arcsec . An alternative is that the background lensing could be produced temporarily (over a few years) by microlensing caused by a star in the foreground galaxy. A very large number of stars in background galaxies will at any one time be in the process of being microlensed by stars in foreground galaxies. This will normally be an undetectable effect; but every so often one of these magnified objects will be a SN precursor star.

\subsection{Conclusions and next steps}

Of the four explanations we have considered for our slow-blue transients, TDEs seem to be ruled out, and extinction events, while we expect them to happen at some level, seem a forced explanation for what we are seeing. 

Microlensing looks promising - it is a phenomenon that {\em must} be happening at some level, the timescale is about right, it explains the slow smooth nature of the majority of light curves, and possibly also the weak broad lines. It is a relatively simple and testable model, and holds the prospect of both accretion disc andBLR mapping. The required models have already been developed for multiple-quasar analysis, and have produced intriguing results \citep{Morgan2010,Blackburne2011,Sluse2012}. On the other hand, accretion instabilities as an explanation has the strong appeal that we need something like this to explain more modest variability in larger numbers of AGN, including objects at very low redshift that cannot plausibly be due to microlensing. On the other hand, in contrast to microlensing, we do not yet have a convincing physical model. It is of course quite possible that we are seeing some objects of each type - microlensing and intrinsic high amplitude variability.

How can we make progress? (i) We need to look for evidence of foreground objects, which is a prediction of the microlensing model. This will probably require deep post-outburst spectra and HST or AO imaging. (ii) We need continued long term monitoring, to construct $\sim 20$ year light curves or even longer. As well as fitting models, the key question is - do they do it again? (iii) We need spectroscopic monitoring. The microlensing model predicts that the broad lines will have light curves that are broader and flatter than the continuum. For intrinsic variability, we need model predictions of how the BLR should respond to such large amplitude changes, as opposed to the more modest changes that have been tracked in reverberation studies. (iv) We need a much larger sample to examine dependence on various parameters - for example luminosity or redshift in the intrinsic case, or lens mass and distance for the microlensing case. In the near term, the TDSS project will deliver additional spectra for hypervariable objects over the entire SDSS and PS1 footprint \citep{Morganson2015}. In the medium term, LSST should produce well sampled light curves of a large number of such extreme variables. (v) We need to catch some objects early in their rise and watch the whole outburst. This can only be done by systematic monitoring of very large numbers of AGN.


\section{Acknowledgements}  \label{sec:acknow}

The Pan-STARRS1 (PS1) Surveys have been made possible through contributions of the Institute for Astronomy, the University of Hawaii, the Pan-STARRS Project Office, the Max-Planck Society and its participating institutes, the Max Planck Institute for Astronomy, Heidelberg and the Max Planck Institute for Extraterrestrial Physics, Garching, The Johns Hopkins University, Durham University, the University of Edinburgh, Queen's University Belfast, the Harvard-Smithsonian Center for Astrophysics, the Las Cumbres Observatory Global Telescope Network Incorporated, the National Central University of Taiwan, the Space Telescope Science Institute, the National Aeronautics and Space Administration under Grant No. NNX08AR22G issued through the Planetary Science Division of the NASA Science Mission Directorate, the National Science Foundation under Grant No. AST-1238877, and the University of Maryland.

The Liverpool Telescope is operated on the island of La Palma by Liverpool John Moores University in the Spanish Observatorio del Roque de los Muchachos of the Instituto de Astrofisica de Canarias with financial support from the UK Science and Technology Facilities Council. 

The William Herschel Telescope is operated on the island of La Palma by the Isaac Newton Group in the Spanish Observatorio del Roque de los Muchachos of the Instituto de Astrof'sica de Canarias.

The research leading to these results has received funding from the European Research Council under the European Union's Seventh Framework Programme (FP7/2007-2013)/ERC Grant agreement n$^{\rm o}$ [291222]  (PI : S. J. Smartt).

MF acknowledges support by the European Union FP7 programme through ERC grant number 320360.

Over an extended period, we have comments from three separate referees, all of which led to some very useful improvements. Peer review is a difficult process, but its always better in the end!



\bibliographystyle{mn2e-williams}
\setlength{\bibhang}{2.0em}
\setlength\labelwidth{0.0em}
\bibliography{slowblue-papers}


\newpage
\appendix

\section{Tables}

Here we provide the full versions of the tables referred to in the text.


\begin{table*}
\caption{Basic sample properties. ``Name'' is a short version of the standard PS1 co-ordinate name, used to cross-link to other tables. }

\begin{tabular}{|l|l|l|l|l|l|}
\hline
Name & Transient ID & RA(2000.0) & Dec (2000.0) &  Flag date & WHT observation date  \\ 

\hline
  J012514 & PS1-12bwl & 01:25:14.09 & +48:05:51.8 & 2012-10-19 & 2014-12-17\\
  J012714 & PS1-12box & 01:27:14.63 & +00:52:24.7 & 2012-10-10 & ---\\
  J025633 & PS1-12bke & 02:56:33.77 & +37:07:12.4 & 2012-09-02 & 2012-09-21\\
  J031240 & PS1-12el & 03:12:40.86 & +18:36:41.1 & 2011-12-21 & 2011-12-21\\
  J033730 & PS1-12bxw & 03:37:30.14 & -07:23:30.2 & 2012-10-25 & ---\\
  J044918 & PS1-12blj & 04:49:18.16 & +11:59:39.5 & 2012-09-19 & 2013-02-11\\
  J061829 & PS1-12et & 06:18:29.11 & +35:35:52.5 & 2012-01-06 & 2013-02-09\\
  J080223 & PS1-12ni & 08:02:23.20 & +28:31:11.8 & 2012-02-21 & ---\\
  J081145 & PS1-12gd & 08:11:45.50 & +15:55:04.9 & 2012-01-30 & ---\\
  J081445 & PS1-12fv & 08:14:45.09 & +23:26:30.4 & 2012-01-28 & 2014-02-07\\
  J081728 & PS1-12fw & 08:17:28.63 & +26:27:20.6 & 2012-01-28 & ---\\
  J081916 & PS1-12fa & 08:19:16.20 & +33:14:05.9 & 2012-01-07 & 2014-02-07\\
  J083544 & PS1-13cu & 08:35:44.41 & +10:08:01.3 & 2013-01-08 & 2013-12-03\\
  J083714 & PS1-12on & 08:37:14.14 & +26:09:32.6 & 2012-02-25 & ---\\
  J084305 & PS1-13jh & 08:43:05.55 & +55:03:51.4 & 2013-02-03 & 2013-02-09\\
  J085220 & PS1-13cl & 08:52:20.12 & +25:57:01.4 & 2013-01-07 & 2013-03-31\\
  J085759 & PS1-13cm & 08:57:59.89 & +25:54:54.5 & 2013-01-07 & 2013-03-12\\
  J090119 & PS1-12mv & 09:01:19.11 & +06:29:43.6 & 2012-02-18 & 2012-03-02\\
  J090244 & PS1-12fc & 09:02:44.51 & +04:52:10.9 & 2012-01-07 & 2013-03-31\\
  J090514 & PS1-12op & 09:05:14.12 & +50:36:28.5 & 2012-02-25 & 2013-02-12\\
  J092358 & PS1-13di & 09:23:58.46 & +62:47:59.6 & 2013-01-12 & ---\\
  J092635 & PS1-12np & 09:26:35.70 & +07:25:32.7 & 2012-02-21 & 2013-02-11\\
  J094309 & PS1-12fl & 09:43:09.96 & +28:35:08.4 & 2012-01-22 & 2014-02-07\\
  J094511 & PS1-12hy & 09:45:11.08 & +17:45:44.8 & 2012-02-07 & 2013-05-15\\
  J094612 & PS1-12fo & 09:46:12.91 & +19:50:28.7 & 2012-01-23 & --- \\
  J102632 & PS1-12cni & 10:26:32.22 & +05:35:08.0 & 2012-12-11 & 2012-12-20\\
  J103511 & PS1-12pa & 10:35:11.67 & +46:04:46.9 & 2012-03-01 & ---\\
  J103726 & PS1-13jo & 10:37:26.93 & -00:38:52.4 & 2013-02-05 & ---\\
  J103837 & PS1-12pb & 10:38:37.10 & +02:11:19.8 & 2012-03-01 & 2013-02-11\\
  J104556 & PS1-12ow & 10:45:56.48 & +05:26:56.2 & 2012-02-25 & 2013-03-30\\
  J104617 & PS1-12qf & 10:46:17.75 & +55:33:36.1 & 2012-03-08 & ---\\
  J105040 & PS1-13ti & 10:50:40.83 & +39:17:35.6 & 2013-02-11 & \\
  J105402 & PS1-12rv & 10:54:02.18 & +16:57:37.8 & 2013-03-17 & ---\\
  J105502 & PS1-13eg & 10:55:02.00 & +33:00:02.5 & 2013-01-20 & 2014-02-07\\
  J110805 & PS1-12yi & 11:08:05.81 & +62:15:00.8 & 2012-04-03 & 2013-02-11\\
  J111547 & PS1-13ty & 11:15:47.78 & +65:20:25.9 & 2013-02-14 & ---\\
  J111706 & PS1-13eh & 11:17:06.68 & -01:02:29.0 & 2013-01-20 & ---\\
  J113309 & PS1-13ud & 11:33:09.68 & -03:39:09.5 & 2013-02-15 & ---\\
  J114742 & PS1-13zi & 11:47:42.78 & +65:05:54.8 & 2013-03-05 & ---\\
  J115553 & PS1-13ch & 11:55:53.06 & +39:36:42.1 & 2012-12-30 & ---\\
  J120240 & PS1-12pg & 12:02:40.91 & +29:50:30.0 & 2012-03-01 & ---\\
  J120921 & PS1-12mp & 12:09:21.46 & +66:53:06.8 & 2012-02-15 & 2012-02-24\\
  J121834 & PS1-12ns & 12:18:34.46 & +06:59:49.8 & 2012-02-21 & ---\\
  J122417 & PS1-12we & 12:24:17.03 & +18:55:29.4 & 2012-03-25 & --- \\
  J124044 & PS1-12fz & 12:40:44.85 & +12:53:21.2 & 2012-01-19 & ---\\
  J124728 & PS1-13aab & 12:47:28.03 & +24:56:53.8 & 2013-03-10 & 2013-05-14\\
  J133004 & PS1-13zt & 13:30:04.98 & +15:22:30.8 & 2013-03-07 & 2013-06-09\\
  J133155 & PS1-12yp & 13:31:55.91 & +23:54:05.7 & 2012-04-09 & ---\\
  J135846 & PS1-12yt & 13:58:46.66 & +61:54:09.1 & 2012-04-02 & 2014-06-26\\
  J141056 & PS1-12yq & 14:10:56.35 & +59:30:31.6 & 2012-04-09 & 2013-03-31\\
  J142232 & PS1-12agr & 14:22:32.45 & +01:40:26.7 & 2012-04-23 & 2013-02-11\\
  J142446 & PS1-12arh & 14:24:46.21 & +46:13:48.7 & 2012-05-23 & 2012-02-05\\
  J142902 & PS1-12apk & 14:29:02.69 & +16:24:29.9 & 2012-05-19 & 2013-02-09\\
  J143531 & PS1-12nc & 14:35:31.51 & +07:13:32.7 & 2012-02-18 & 2013-02-09\\
  J145240 & PS1-12nf & 14:52:40.70 & +06:39:31.6 & 2012-02-18 & ---\\
  J150042 & PS1-12agw & 15:00:42.64 & +52:42:38.5 & 2012-04-23 & 2014-07-24\\
  J150210 & PS1-12apg & 15:02:10.46 & +23:09:15.3 & 2012-05-16 & 2013-04-30\\
  J151201 & PS1-12ajx & 15:12:01.72 & +05:00:56.2 & 2012-05-04 & 2013-05-15\\
  J151944 & PS1-12aiu & 15:19:44.00 & +00:11:47.4 & 2012-04-28 & 2014-06-26\\
  J154445 & PS1-12ars & 15:44:45.52 & +27:29:14.4 & 2012-05-26 & 2013-04-30\\
  J154513 & PS1-12bjg & 15:45:13.66 & +27:50:19.1 & 2012-08-23 & ---\\
  J154950 & none & 15:49:50.69 & +14:49:30.0 & 2011-06-03 & 2011-06-09\\
  J155427 & none & 15:54:27.15 & +52:35:13.9 & 2011-05-09 & 2014-06-24\\
\hline\end{tabular}
\end{table*}

\newpage
\begin{table*}
\setcounter{table}{0}
\caption{Continued}
\begin{tabular}{l l l l l l l l}
Name & Transient ID &  RA(2000.0) & Dec (2000.0) & Flag date & WHT observation date  \\ 
\hline

 J160329 & PS1-12aha & 16:03:29.42 & +06:05:05.8 & 2012-04-23 & 2013-05-14\\
  J160332 & PS1-12atz & 16:03:32.98 & +58:03:05.9 & 2012-06-06 & 2013-05-15\\
  J161022 & PS1-12aji & 16:10:22.86 & +08:38:46.2 & 2012-04-28 & 2013-08-08\\
  J170800 & PS1-12arz & 17:08:00.75 & +10:24:25.0 & 2012-05-26 & ---\\
  J170845 & PS1-12asa & 17:08:45.13 & +19:05:11.7 & 2012-05-26 & 2013-06-09\\
  J172534 & PS1-12axc & 17:25:34.88 & +08:35:45.6 & 2012-06-14 & ---\\
  J172639 & PS1-12apd & 17:26:39.90 & +61:27:06.5 & 2012-05-14 & ---\\
  J175610 & PS1-12bcb & 17:56:10.00 & +46:39:58.7 & 2012-08-09 & 2014-07-23\\
  J202823 & PS1-12axd & 20:28:23.49 & +60:02:33.9 & 2012-06-14 & 2013-08-08\\
  J221241 & PS1-12baa & 22:12:41.55 & +00:30:43.1 & 2012-08-23 & --- \\
  J223210 & PS1-12bjx & 22:32:10.51 & -08:06:21.2 & 2012-08-30 & 2013-06-10\\
  J233237 & PS1-12bzm & 23:32:37.52 & -10:04:44.0 & 2012-11-02 & 2014-07-22\\
  J234953 & PS1-12baj & 23:49:53.52 & -09:16:06.9 & 2012-07-30 & 2014-07-22\\
\end{tabular}
\end{table*}

\newpage
\begin{table*}
\caption{Cross-identifications for the sample. Note that for UKIDSS, ``N/A'' means that the object is outside the UKIDSS footprint, whereas ``-'' means the object is inside the UKIDSS footprint but not seen.}

\begin{tabular}{|l|l|l|l|l|}
\hline
Name & SDSS ID & UKIDSS ID & NVSS ID & CRTS ID \\
\hline
  J012514 & J012514.10+480551.9 & N/A & - & -\\
  J012714 & J012714.65+005224.6 & - & - & CSS121013:012715+005224\\
  J025633 & J025633.76+370712.3 & N/A & 219702 & -\\
  J031240 & J031240.88+183641.1 & N/A & - & -\\
  J033730 & J033730.15-072330.2 & N/A & - & -\\
  J044918 & J044918.15+115939.5 & N/A & - & MLS121114:044918+115940\\
  J061829 & J061829.10+353552.5 & N/A & - & -\\
  J080223 & J080223.16+283111.5 & 433796865918 & - & -\\
  J081145 & J081145.50+155504.9 & N/A & - & -\\
  J081445 & J081445.09+232630.3 & 433802373716 & - & MLS120127:081146+155505\\
  J081728 & J081728.61+262720.8 & - & - & -\\
  J081916 & J081916.20+331405.9 & N/A & - & -\\
  J083544 & J083544.40+100801.2 & - & - & -\\
  J083714 & J083714.13+260932.4 & 433799630267 & - & -\\
  J084305 & J084305.50+550351.0 & N/A & - & -\\
  J085220 & J085220.13+255701.2 & 433799627553 & - & -\\
  J085759 & J085759.89+255454.3 & - & - & CSS121115:085800+255454\\
  J090119 & J090119.10+062943.6 & 433832907807 & - & -\\
  J090244 & J090244.50+045210.9 & 433840230630 & 678186 & -\\
  J090514 & J090514.12+503628.4 & N/A & - & -\\
  J092358 & J092358.39+624759.7 & N/A & - & CSS130108:092358+624800\\
  J092635 & J092635.71+072532.5 & 433877433237 & - & -\\
  J094309 & J094309.96+283508.4 & N/A & 728422 & -\\
  J094511 & J094511.08+174544.7 & N/A & - & CSS111231:094511+174545\\
  J094612 & J094612.91+195028.6 & N/A & - & CSS120121:094613+195028\\
  J102632 & J102632.22+053508.1 & - & - & CSS121114:102632+053508\\
  J103511 & J103511.66+460446.8 & N/A & - & -\\
  J103726 & J103726.92-003852.6 & 433868153861 & - & -\\
  J103837 & J103837.08+021119.7 & 433850380130 & - & -\\
  J104556 & J104556.46+052655.9 & 433837504016 & - & -\\
  J104617 & J104617.71+553336.4 & N/A & - & -\\
  J105040 & J105040.82+391735.6 & N/A & - & -\\
  J105402 & J105402.18+165738.0 & N/A & - & -\\
  J105502 & J105502.00+330002.4 & N/A & - & CSS130105:105502+330004\\
  J110805 & J110805.80+621500.9 & N/A & - & -\\
  J111547 & J111547.76+652025.7 & N/A & - & -\\
  J111706 & J111706.70-010228.8 & - & - & MLS130122:111707-010229\\
  J113309 & J113309.67-033909.6 & 433883390451 & - & -\\
  J114742 & J114742.76+650554.7 & N/A & - & -\\
  J115553 & J115553.04+393642.1 & N/A & - & -\\
  J120240 & J120240.90+295029.9 & N/A & - & -\\
  J120921 & J120921.45+665306.3 & N/A & - & -\\
  J121834 & J121834.46+065949.9 & 433830393521 & 921141 & CSS110104:121834+065950\\
  J122417 & J122417.03+185529.4 & N/A & - & CSS120328:122417+185529\\
  J124044 & J124044.82+125321.5 & - & - & CSS120125:124045+125321\\
  J124728 & J124728.01+245653.6 & 433800999723 & - & -\\
  J133004 & J133004.98+152230.6 & 433805594397 & - & -\\
  J133155 & J133155.90+235405.8 & 433801910070 & - & CSS120301:133156+235405\\
  J135846 & J135846.65+615409.2 & N/A & - & -\\
  J141056 & J141056.34+593031.8 & N/A & - & -\\
  J142232 & J142232.45+014026.8 & 433852135933 & - & -\\
  J142446 & J142446.21+461348.6 & N/A & - & -\\
  J142902 & J142902.67+162429.7 & N/A & - & -\\
  J143531 & J143531.52+071332.6 & 433830056985 & - & -\\
  J145240 & J145240.65+063931.4 & - & - & -\\
  J150042 & J150042.63+524238.5 & N/A & - & -\\
  J150210 & J150210.47+230915.2 & N/A & - & CSS120514:150211+230915\\
  J151201 & J151201.71+050056.1 & 433839619010 & - & -\\
  J151944 & J151944.00+001147.4 & 433862112851 & - & -\\
  J154445 & J154445.51+272914.4 & 433879848087 & - & -\\
  J154513 & J154513.66+275019.0 & 433879854735 & - & -\\
  J154950 & J154950.71+144929.9 & N/A & - & -\\
  J155427 & J155427.16+523513.8 & N/A & - & -\\
\hline
\end{tabular}
\end{table*}

\newpage
\begin{table*}
\setcounter{table}{1}
\caption{Continued}
\begin{tabular}{|l|l|l|l|l|}
\hline
Name & SDSS ID & UKIDSS ID & NVSS ID & CRTS ID \\
\hline
 J160329 & J160329.43+060505.8 & N/A & 1199286 & -\\
  J160332 & J160332.96+580305.8 & N/A & - & -\\
  J161022 & J161022.87+083846.1 & N/A & - & -\\
  J170800 & J170800.74+102425.4 & N/A & - & -\\
  J170845 & J170845.12+190511.7 & N/A & - & -\\
  J172534 & J172534.87+083545.5 & N/A & 1303932 & -\\
  J172639 & J172639.90+612706.7 & N/A & - & -\\
  J175610 & J175609.99+463958.6 & N/A & - & -\\
  J202823 & J202823.50+600234.2 & N/A & - & -\\
  J221241 & J221241.53+003042.7 & 433858000660 & - & CSS120825:221242+003043\\
  J223210 & J223210.51-080621.3 & N/A & - & -\\
  J233237 & J233237.53-100444.1 & N/A & - & -\\
  J234953 & J234953.52-091607.1 & N/A & - & -\\

\hline
\end{tabular}
\end{table*}

\newpage
\begin{table*}
\caption{Host and transient photometry. The host galaxy photometry uses the {\em cmodel} magnitudes from SDSS DR7. The transient magnitudes are 2$^{\prime\prime}$ aperture photometry from the Liverpool Telescope data.}
\begin{tabular}{cccccccc}
\hline
& 
\multicolumn{4}{c}{host galaxy} & 
\multicolumn{3}{c}{transient} \\
Name & g & u-g & g-r & r-i & g$_{max}$ & u-g & g-r \\
\hline
  J012514 & 23.68(0.27) & 0.21(1.00) & 1.02(0.32) & 0.07(0.35) & 19.02 & 0.18(0.04) & 0.10(0.03)\\
  J012714 & 22.76(0.19) & 0.38(0.77) & -0.13(0.34) & 0.92(0.36) & 20.19 & 1.25(0.12) & 0.56(0.05)\\
  J025633 & 21.16(0.04) & 1.01(1.19) & 1.07(0.09) & 0.05(0.09) & 18.6 & 0.51(0.03) & 0.47(0.02)\\
  J031240 & 21.49(0.05) & 0.63(0.25) & 0.20(0.08) & 0.31(0.10) & 19.24 & 0.26(0.05) & -0.05(0.03)\\
  J033730 & 21.37(0.06) & 2.67(1.44) & 1.04(0.07) & 0.42(0.06) & 19.68 & 0.51(0.14) & 0.61(0.06)\\
  J044918 & 22.46(0.10) & 0.49(0.51) & 0.38(0.16) & 0.33(0.20) & 19.26 & 0.54(0.05) & 0.29(0.03)\\
  J061829 & 21.72(0.05) & 0.59(0.23) & 0.38(0.07) & 0.66(0.07) & 16.52 & 0.39(0.01) & 0.29(0.01)\\
  J080223 & 21.85(0.07) & 0.86(0.39) & 0.51(0.09) & 0.33(0.09) & 19.49 & $>$1.81 & 0.91(0.07)\\
  J081145 & 22.37(0.14) & 0.48(0.64) & 0.53(0.18) & -0.22(0.24) & 19.43 & 0.98(0.22) & 0.48(0.08)\\
  J081445 & 22.16(0.08) & 0.13(0.24) & 0.58(0.11) & -0.19(0.13) & 19.71 & 0.04(0.04) & 0.33(0.03)\\
  J081728 & 22.54(0.15) & 1.11(1.11) & 0.61(0.20) & -0.06(0.24) & 20.38 & 1.21(0.30) & 0.46(0.05)\\
  J081916 & 21.49(0.06) & 0.25(0.22) & 0.99(0.07) & 0.67(0.05) & 19.68 & 0.32(0.04) & 0.19(0.03)\\
  J083544 & 21.46(0.04) & -0.03(0.11) & 0.88(0.05) & -0.28(0.06) & 19.26 & -0.25(0.04) & 0.42(0.03)\\
  J083714 & 21.83(0.07) & -0.28(0.16) & 0.74(0.09) & 0.12(0.09) & 20.02 & -0.36(0.05) & 0.34(0.04)\\
  J084305 & 22.25(0.19) & 0.19(0.65) & 0.74(0.24) & 0.65(0.20) & 19.97 & 0.06(0.04) & 0.20(0.03)\\
  J085220 & 20.87(0.03) & 0.44(0.09) & 0.17(0.04) & 0.33(0.04) & 19.5 & 0.22(0.05) & 0.27(0.05)\\
  J085759 & 21.17(0.05) & 0.76(0.21) & -0.02(0.07) & 0.24(0.08) & 19.53 & 0.37(0.04) & 0.19(0.03)\\
  J090119 & 21.50(0.05) & 0.77(0.27) & 0.16(0.09) & 0.41(0.13) & 19.62 & 0.90(0.09) & 0.45(0.04)\\
  J090244 & 21.19(0.04) & 0.23(0.16) & 0.73(0.06) & 0.64(0.05) & 19.54 & 0.33(0.05) & 0.21(0.03)\\
  J090514 & 22.31(0.18) & 0.01(0.39) & 1.15(0.19) & -0.64(0.25v & 19.82 & -0.36(0.04) & 0.28(0.03)\\
  J092358 & 21.20(0.06) & 1.80(0.65) & 0.37(0.10) & 0.32(0.12) & 19.87 & 0.47(0.12) & 0.51(0.04)\\
  J092635 & 21.43(0.06) & 1.83(0.90) & 1.16(0.07) & 0.73(0.05) & 19.76 & 0.27(0.06) & 0.25(0.05)\\
  J094309 & 20.45(0.02) & 0.10(0.06) & 0.93(0.03) & -0.02(0.03) & 18.84 & 0.19(0.03) & 0.36(0.02)\\
  J094511 & 22.31(0.10) & 0.89(0.57) & 0.61(0.13) & 0.40(0.13) & 19.94 & 0.01(0.06) & 0.16(0.05)\\
  J094612 & 22.12(0.08) & 1.37(0.67) & 0.65(0.11) & -0.20(0.14) & 18.33 & 1.05(0.38) & 0.73(0.05)\\
  J102632 & 21.65(0.12) & 1.35(0.77) & 0.15(0.18) & 0.41(0.18) & 20.14 & 1.59(0.21) & 1.05(0.04)\\
  J103511 & 21.80(0.08) & 1.56(0.95) & 0.34(0.11) & 0.21(0.12) & 19.82 & 0.23(0.05) & 0.20(0.03)\\
  J103726 & 22.61(0.15) & 1.62(1.49) & 0.83(0.18) & 0.16(0.17) & 20.11 & 0.21(0.05) & 0.42(0.02)\\
  J103837 & 21.46(0.06) & 1.04(0.31) & 0.16(0.09) & 0.53(0.09) & 19.51 & 0.49(0.05) & 0.09(0.03)\\
  J104556 & 21.40(0.05) & 0.13(0.18) & 0.38(0.07) & 0.29(0.08) & 19.21 & 0.19(0.05) & 0.35(0.03)\\
  J104617 & 21.59(0.05) & 1.17(0.49) & 0.69(0.07) & 0.21(0.08) & 19.99 & 1.05(0.11) & 0.62(0.04)\\
  J105040 & 20.44(0.03) & 0.30(0.09) & 1.02(0.04) & 0.27(0.03) & 18.84 & 0.05(0.02) & 0.30(0.02)\\
  J105402 & 22.03(0.11) & 0.62(0.48) & 0.74(0.14) & 0.21(0.16) & 19.53 & $>$1.1 & 0.99(0.07)\\
  J105502 & 20.46(0.03) & 0.39(0.10) & 0.87(0.04) & 0.52(0.03) & 18.96 & 0.15(0.02) & 0.06(0.02)\\
  J110805 & 21.90(0.09) & 1.32(0.82) & 0.93(0.11) & 0.71(0.08) & 19.27 & 0.04(0.03) & 0.01(0.03)\\
  J111547 & 20.45(0.03) & 0.55(0.13) & 0.58(0.04) & 0.55(0.04) & 18.93 & 0.13(0.02) & 0.09(0.02)\\
  J111706 & 22.29(0.13) & 0.28(0.33) & 0.73(0.16) & 0.43(0.15) & 19.41 & 0.20(0.35) & 0.75(0.15)\\
  J113309 & 20.58(0.05) & 0.46(0.16) & 0.79(0.06) & 0.66(0.04) & 18.92 & 0.14(0.02) & 0.12(0.02)\\
  J114742 & 21.83(0.07) & 1.44(0.71) & 0.19(0.11) & 0.17(0.14) & 20.2 & 0.61(0.06) & 0.34(0.02)\\
  J115553 & 21.92(0.11) & 0.59(0.43) & 0.41(0.15) & 0.87(0.12) & 19.7 & 0.10(0.04) & 0.12(0.03)\\
  J120240 & 23.00(0.19) & -0.52(0.31) & 0.68(0.23) & -0.48(0.31) & 22.77 & -0.53(0.22) & 0.36(0.25)\\
  J120921 & 22.24(0.14) & 0.98(1.07) & 0.84(0.18) & -0.55(0.31) & 20.24 & 1.49(0.20) & 0.81(0.03)\\
  J121834 & 22.38(0.12) & 1.77(1.51) & 0.31(0.18) & 0.41(0.18) & 20.53 & 0.33(0.15) & 0.05(0.11)\\
  J122417 & 19.67(0.03) & 0.74(0.11) & 0.23(0.04) & 0.10(0.06) & 18.1 & 0.82(0.04) & 0.74(0.01)\\
  J124044 & 21.31(0.06) & 0.50(0.23) & 0.08(0.10) & 0.37(0.13) & 20.55 & 1.30(0.17) & 0.95(0.04)\\
  J124728 & 20.73(0.03) & 0.71(0.20) & 0.95(0.04) & 0.49(0.03) & 18.95 & 0.02(0.02) & 0.07(0.01)\\
  J133004 & 21.37(0.05) & 0.58(0.24) & 0.74(0.06) & 0.56(0.06) & 18.97 & 0.22(0.03) & 0.19(0.02)\\
  J133155 & 22.58(0.13) & 0.39(0.47) & 1.36(0.14) & 0.95(0.07) & 19.87 & -0.54(0.05) & 0.05(0.06)\\
  J135846 & 21.71(0.11) & 0.20(0.29) & 0.43(0.15) & 0.78(0.12) & 19.88 & 0.08(0.05) & 0.17(0.03)\\
  J141056 & 20.62(0.03) & 0.78(0.15) & 0.31(0.04) & 0.39(0.04) & 18.77 & 0.37(0.02) & 0.14(0.02)\\
  J142232 & 23.61(0.36) & 1.07(1.69) & 1.48(0.39) & 0.06(0.26) & 19.74 & 0.01(0.04) & 0.22(0.03)\\
  J142446 & 19.31(0.01) & 0.36(0.04) & 0.15(0.01) & 0.45(0.01) & 16.74 & 0.52(0.02) & 0.14(0.02)\\
  J142902 & 20.78(0.03) & 0.53(0.16) & 0.56(0.04) & 0.60(0.04) & 19.22 & 0.25(0.02) & 0.09(0.02)\\
  J143531 & 21.51(0.07) & 0.20(0.27) & 0.54(0.10) & 0.25(0.11) & 20.15 & 0.06(0.06 & 0.24(0.05)\\
  J145240 & 23.51(0.29) & -0.23(0.60) & 0.62(0.38) & 0.91(0.31) & 22.3 & ---  & 0.60(0.83)\\
  J150042 & 22.12(0.09) & 0.72(0.45) & 0.16(0.13) & 0.36(0.16) & 19.9 & 0.31(0.05) & 0.21(0.03)\\
  J150210 & 21.48(0.06) & 0.38(0.27) & 0.43(0.08) & 0.54(0.07) & 18.82 & -0.42(0.02) & -0.16(0.02)\\
  J151201 & 21.87(0.06) & 0.33(0.24) & 0.17(0.09) & 0.09(0.11) & 19.92 & 0.15(0.05) & 0.17(0.03)\\
  J151944 & 21.42(0.04) & 0.59(0.17) & 0.37(0.06) & 0.66(0.06) & 19.15 & 0.33(0.02) & -0.02(0.02)\\
  J154445 & 20.98(0.03) & 0.40(0.12) & 0.19(0.05) & 0.33(0.06) & 19.05 & 0.33(0.02) & -0.05(0.02)\\
  J154513 & 21.31(0.04) & 0.71(0.22) & 0.36(0.06) & 0.17(0.07) & 21.24 & 0.56(0.17) & 0.13(0.07)\\
  J154950 & 21.82(0.07) & 0.77(0.32) & 0.52(0.09) & 0.11(0.10) & 18.95 & 0.27(0.73) & 0.27(0.12)\\
  J155427 & 22.13(0.10) & 0.35(0.38) & 0.54(0.13) & 0.79(0.12) & 19.33 & 0.33(0.05) & 0.07(0.04)\\
\end{tabular}
\end{table*}

\newpage
\begin{table*}
\setcounter{table}{2}
\caption{Continued}
\begin{tabular}{cccccccc}
\hline
& 
\multicolumn{4}{c}{host galaxy} & 
\multicolumn{3}{c}{transient} \\
Name & g & u-g & g-r & r-i & g$_{max}$ & u-g & g-r \\
\hline
  J160329 & 22.61(0.11) & -0.32(0.24) & 0.83(0.14) & -0.01(0.15) & 19.84 & -0.19(0.04) & 0.29(0.02)\\
  J160332 & 22.07(0.08) & 0.53(0.38) & 0.56(0.11) & 0.38(0.10) & 20.19 & 0.00(0.05) & 0.46(0.03)\\
  J161022 & 21.33(0.05) & 0.59(0.22) & 0.22(0.07) & 0.42(0.07) & 19.97 & 0.38(0.04) & 0.17(0.03)\\
  J170800 & 22.52(0.15) & 0.26(0.57) & 0.47(0.21) & 0.06(0.27) & 19.98 & 0.99(0.11) & 0.60(0.14)\\
  J170845 & 21.45(0.05) & 1.31(0.36) & 0.43(0.07) & 0.90(0.06) & 19.92 & 0.54(0.05) & -0.05(0.23)\\
  J172534 & 23.58(0.40) & 1.35(1.62) & 1.44(0.43) & 0.29(0.26) & 21.01 & 0.65(0.10) & 0.59(0.03)\\
  J172639 & 22.65(0.21) & 0.31(0.79) & 0.71(0.27) & -0.02(0.29) & 22.51 & -0.50(0.44) & 0.56(0.31)\\
  J175610 & 21.19(0.05) & 0.99(0.33) & 0.13(0.08) & 0.40(0.08) & 19.51 & 0.51(0.07) & 0.06(0.04)\\
  J202823 & 22.97(0.17) & 1.10(0.98) & 0.79(0.21) & 0.61(0.18) & 18.92 & 0.89(0.05) & 0.61(0.02)\\
  J221241 & 18.97(0.02) & 0.19(0.05) & 0.08(0.03) & -0.04(0.04) & 17.43 & 1.84(0.03) & 0.89(0.01)\\
  J223210 & 20.09(0.03) & -0.32(0.06) & 0.90(0.04) & 0.24(0.03) & 18.29 & -0.13(0.02) & 0.15(0.02)\\
  J233237 & 21.93(0.09) & -0.43(0.19) & 0.50(0.12) & 0.39(0.12) & 20.02 & -0.25(0.04) & 0.31(0.03)\\
  J234953 & 21.65(0.07) & -0.43(0.15) & 0.79(0.09) & 0.07(0.11) & 20.02 & -0.19(0.05) & 0.28(0.04)\\
\hline
\end{tabular}
\end{table*}

\newpage
\begin{table*}
\caption{Various quantities derived from the light curves, spectra. (1) Usual short name (2) Colour type as defined in the text - 1=red, 2=blue, 3=ultra-blue. (3) Transient amplitude in g-band - SDSS magnitude minus PS1/LT magnitude at time of flag. The quantity in brackets is the error. (4) Spectral type. (5) Spectroscopic redshift. (6) Photometric redshift, from SDSS DR7. The quantity in brackets is the error. (7) Early decay rate in mags/month, as defined in the text. The first quantity in brackets is the error; the second in brackets is the slope in the AGN rest-frame, i.e. multiplied by $1=z$.  (8) Notes. ``DR7'' and ``DR9'' refer to SDSS releases. Objects are morphologically classified as galaxy in both DR7 and DR9 except where noted. 
}
\begin{tabular}{|l|c|l|l|l|l|l|l|}
\hline
  \multicolumn{1}{|c|}{name} &
  \multicolumn{1}{c|}{ctype} &
  \multicolumn{1}{c|}{g-amp} &
  \multicolumn{1}{c|}{spectype} &
  \multicolumn{1}{c|}{specz} &
  \multicolumn{1}{c|}{photz} &
  \multicolumn{1}{c|}{slope} &
  \multicolumn{1}{c|}{note} 
   \\
\hline

  J012514 & 2 & 4.66(0.27) & vstar  & 0.0   & 0.077(0.077)  & +4.4300(0.7700)              & \\
  J012714 & 1 & 2.57(0.19) & nospec & --    & 0.090(0.009)  & +0.3300(0.0300)             & \\
  J025633 & 1 & 2.56(0.04) & smooth & --    & 0.123(0.118)  & +0.3800(0.0600)              & DR9 class=star\\
  J031240 & 2 & 2.25(0.05) & AGN    & 0.891 & 0.398(0.151)  & +0.0157(0.0063) [+0.0297] &  \\
  J033730 & 1 & 1.69(0.06) & nospec & --    & 0.121(0.022)  & +0.4110(0.0350)            &  \\
  J044918 & 1 & 3.20(0.10) & vstar  & 0.0   & 0.362(0.028)  & -0.2407(0.3160)           &  \\
  J061829 & 2 & 5.20(0.05) & vstar  & 0.0   & --            & +0.0530(0.0100)            & DR7 class=star \\
  J080223 & 1 & 2.36(0.07) & nospec & --    & 0.143(0.090)  & +1.1330(0.1750)            &  \\
  J081145 & 1 & 2.94(0.14) & nospec & --    & 0.163(0.105)  & +1.2890(0.1610)            &  \\
  J081445 & 2 & 2.45(0.08) & AGN    & 1.17  & 0.286(0.085)  & +0.0169(0.0050)  [+0.0367] &  \\
  J081728 & 1 & 2.16(0.15) & nospec & --    & 0.039(0.020)  & +0.5000(0.0490)            &  \\
  J081916 & 2 & 1.81(0.06) & AGN    & 0.426 & 0.432(0.116)  & +0.0502(0.0061) [+0.0716] & also spec from SDSS DR7; ROSAT target \\
  J083544 & 3 & 2.20(0.04) & AGN    & 1.327 & --            & -0.0244(0.0066) [-0.0568] & DR7 class=star\\
  J083714 & 3 & 1.81(0.07) & nospec & --    & 0.215(0.077)  & +0.0522(0.0031)            &  \\
  J084305 & 2 & 2.28(0.19) & AGN    & 0.894 & 0.490(0.078)  & +0.0061(0.0066) [+0.0116]  & \\
  J085220 & 2 & 1.37(0.03) & AGN    & 0.854 & 0.442(0.021)  & +0.0644(0.0280) [+0.1194]  & \\
  J085759 & 2 & 1.64(0.05) & AGN    & 0.746 & 0.908(0.320)  & +0.0488(0.0290) [+0.0852]  & \\
  J090119 & 1 & 1.88(0.05) & SNIIn  & 0.11  & 0.300(0.195)  & +0.2783(0.0398) [+0.3089]  & \\
  J090244 & 2 & 1.65(0.04) & AGN    & 0.437 & 0.458(0.081)  & +0.0207(0.0091) [+0.0297]  & \\
  J090514 & 3 & 2.49(0.18) & AGN    & 1.29  & 0.236(0.178)  & +0.0266(0.0025) [0.0609]   & \\
  J092358 & 1 & 1.33(0.06) & nospec & --    & 0.061(0.022)  & +0.4540(0.0412)            & \\
  J092635 & 2 & 1.67(0.06) & AGN    & 0.465 & 0.137(0.039)  & +0.0043(0.0073) [+0.0063]   & \\
  J094309 & 2 & 1.61(0.02) & AGN    & 1.269 & 0.181(0.119)  & -0.0039(0.0021) [-0.0088]  & DR9 class=star\\
  J094511 & 2 & 2.37(0.10) & AGN    & 0.758 & 0.149(0.041)  & +0.0486(0.0036) [+0.085]   & \\
  J094612 & 1 & 3.79(0.08) & SNIc   & 0.175 & 0.046(0.047)  & +1.4456(0.0939) [+1.6986]   & spec from NOT; SN2012il\\
  J102632 & 1 & 1.51(0.12) & SNIIp  & 0.045 & 0.055(0.027)  & +1.1521(0.1059) [+1.2039]  & \\
  J103511 & 2 & 1.98(0.08) & nospec & --    & 0.039(0.018)  & +0.0285(0.0092)            & \\
  J103726 & 2 & 2.50(0.15) & nospec & --    & 0.086(0.037)  & +0.1150(0.0400)              & \\
  J103837 & 2 & 1.95(0.06) & AGN    & 0.62  & 0.061(0.020)  & +0.0226(0.0079) [+0.0366]   & \\
  J104556 & 2 & 2.19(0.05) & AGN    & 0.995 & 0.321(0.042)  & -0.0363(0.0019) [-0.0724]  & DR9 class=star\\ 
  J104617 & 1 & 1.60(0.05) & nospec & --    & 0.075(0.056)  & +0.1191(0.0094)            & \\
  J105040 & 2 & 1.60(0.03) & AGN    & 0.306 & 0.302(0.043)  & +0.0053(0.0244) [+0.0069]  & spec from SDSS DR7; ROSAT target\\
  J105402 & 1 & 2.50(0.11) & nospec & --    & 0.329(0.033)  & +1.1183(0.2619)            & \\
  J105502 & 2 & 1.50(0.03) & AGN    & 0.417 & 0.337(0.068)  & +0.0127(0.0144) [+0.0180]  & \\
  J110805 & 2 & 2.63(0.09) & AGN    & 0.536 & 0.132(0.044)  & +0.0419(0.0032) [+0.0644]   & \\
  J111547 & 2 & 1.52(0.03) & nospec & --    & 0.459(0.146)  & +0.1787(0.0053)            & \\
  J111706 & 2 & 2.88(0.13) & nospec & --    & 0.304(0.033)  & +1.2800(0.0760)            & \\
  J113309 & 2 & 1.66(0.05) & nospec & --    & 0.489(0.118)  & -0.0459(0.0082)            & \\
  J114742 & 1 & 1.63(0.07) & nospec & --    & 0.100(0.049)  & -0.0001(0.0076) [+0.0882]   & DR9 class=star \\
  J115553 & 2 & 2.22(0.11) & nospec & --    & 0.488(0.183)  & +0.0683(0.0081)            &  \\
  J120240 & 3 & 0.23(0.19) & nospec & --    & 0.159(0.145)  & +0.0551(0.0344)            & \\ 
  J120921 & 1 & 2.00(0.14) & SNIa   & 0.058 & 0.072(0.090)  & +0.3616(0.0253) [+0.3823]   & \\
  J121834 & 2 & 1.85(0.12) & nospec & --    & 0.093(0.050)  & +0.0275(0.0059)            &  \\
  J122417 & 1 & 1.57(0.03) & SNII   & 0.019 & 0.021(0.006)  & +0.5816(0.0485) [0.5927]   & spec from INT \\
  J124044 & 1 & 0.76(0.06) & nospec & --    & 0.315(0.070)  & +0.0896(0.0144)            &  \\
  J124728 & 2 & 1.78(0.03) & AGN    & 0.454 & 0.305(0.061)  & -0.0155(0.0041) [-0.0225]  &  \\
  J133004 & 2 & 2.40(0.05) & AGN    & 0.357 & 0.412(0.114)  & -0.0420(0.0161) [-0.0570]  &  \\
  J133155 & 3 & 2.71(0.13) & nospec & --    & 0.581(0.065)  & +0.1537(0.0115)            & \\
  J135846 & 2 & 1.83(0.11) & AGN    & 0.845 & 0.584(0.094)  & +0.0084(0.0018) [+0.0155]   &  \\
  J141056 & 2 & 1.85(0.03) & AGN    & 0.674 & 0.052(0.063)  & +0.0452(0.0057) [+0.0757]  &  \\
  J142232 & 2 & 3.87(0.36) & AGN    & 1.079 & 0.200(0.113)  & +0.0201(0.0058) [+0.0418]   &  \\
  J142446 & 2 & 2.57(0.01) & SNIc   & 0.107 & 0.110(0.022)  & +0.3342(0.0046) [+0.3700]   & PTF12dam\\ 
  J142902 & 2 & 1.56(0.03) & AGN    & 0.439 & 0.422(0.168)  & +0.0721(0.0040) [+0.1038]  &  \\
  J143531 & 2 & 1.36(0.07) & AGN    & 0.439 & 0.098(0.059)  & +0.0310(0.0029) [+0.0446]  & DR9 class=star \\
  J145240 & --& 1.21(0.29) & nospec & --    & 0.454(0.139)  & +0.2122(0.1030)            & \\
  J150042 & 2 & 2.22(0.09) & AGN    & 0.752 & 1.145(0.202)  & +0.0672(0.0037) [+0.1178]  &  \\
  J150210 & 3 & 2.66(0.06) & AGN    & 0.630 & 0.438(0.078)  & -0.0175(0.0085) [-0.0285]  &  \\
  J151201 & 2 & 1.95(0.06) & AGN    & 0.933 & 0.138(0.053)  & -0.0075(0.0037) [-0.0145]  & \\
  J151944 & 2 & 2.27(0.04) & AGN    & 0.534 & 0.443(0.191)  & +0.0114(0.0039) [+0.0175]  &  \\
  J154445 & 2 & 1.93(0.03) & AGN    & 0.548 & 0.139(0.024)  & +0.0039(0.0050) [+0.0060]  &  \\
 
 \end{tabular}
\end{table*}

\newpage
\begin{table*}
\setcounter{table}{3}
\caption{Continued}

\begin{tabular}{|l|c|l|l|l|l|l|l|}
Name & ctype & g-amp & spec type & specz  & photz & slope  & note\\
\hline
  J154513 & 2 & 0.07(0.04) & nospec & --    & 0.096(0.038)  & +0.0311(0.0517)            & \\
  J154950 & 2 & 2.87(0.07) & SNIa   & 0.12  & 0.146(0.092)  & +0.2717(0.017) [+0.3043]   & SN2011er\\
  J155427 & 2 & 2.78(0.10) & AGN    & 0.572 & 0.536(0.147)  & -0.0180(0.0044) [-0.0283]  &  \\
  J160329 & 3 & 2.77(0.11) & AGN    & 1.412 & 0.233(0.070)  & +0.0058(0.0259) [+0.0140]  & \\
  J160332 & 2 & 1.88(0.08) & AGN    & 1.044 & 0.237(0.084)  & +0.0233(0.0078) [+0.0476]  & \\
  J161022 & 2 & 1.36(0.05) & AGN    & 1.986 & 0.110(0.022)  & +0.1504(0.0784) [+0.449]   & DR9 class=star \\
  J170800 & 1 & 2.54(0.15) & nospec & --    & 0.118(0.061)  & +1.0754(0.1826)            & \\
  J170845 & 2 & 1.53(0.05) & AGN    & 0.586 & 0.133(0.056)  & -0.0084(0.0111) [-0.0133]  &  \\
  J172534 & 1 & 2.57(0.40) & nospec & --    & 0.222(0.100)  & +0.2364(0.1211)            &  \\
  J172639 & 3 & 0.4(0.21)  & nospec & --    & 0.299(0.048)  & +0.2273(0.1496)            &  \\
  J175610 & 2 & 1.68(0.05) & AGN    & 0.677 & 0.049(0.013)  & -0.0012(0.0053) [-0.0020]  &  \\
  J202823 & 1 & 4.05(0.17) & vstar  & 0.0   & 0.152(0.076)  & +0.1928(0.0365)            &  \\
  J221241 & 1 & 1.54(0.02) & SNI-pec& 0.0137& 0.185(0.135)  & +0.3104(0.0283) [+0.3147]  & PTF12gzk;DR9-spec-starburst\\
  J223210 & 3 & 1.80(0.03) & AGN    & 0.276 & 0.269(0.064)  & -0.0054(0.0015) [-0.0069]  &  \\
  J233237 & 3 & 1.91(0.09) & AGN    & 1.471 & 0.257(0.050)  & +0.0113(0.0052) [+0.0279]  & DR9 class=star \\
  J234953 & 3 & 1.63(0.07) & AGN    & 1.278 & 0.586(0.293)  & +0.0017(0.0338) [+0.0039]  & DR9 class=star \\

\hline\end{tabular}
\end{table*}

\newpage
\begin{table*}
\caption{Quantities measured from the WHT spectra. Fluxes are in units of $10^{-15}$ erg cm$^{-2}$ s$^{-1}$. Equivalent widths are AGN rest frame equivalent widths in units of \AA .
}
\begin{tabular}{|l|r|l|r|r|l|l|l|l|l|l|}
\hline
  \multicolumn{1}{|c|}{name} &
  \multicolumn{1}{c|}{z} &
  \multicolumn{1}{c|}{F(MgII)} &
  \multicolumn{1}{c|}{F(MgII$_{cont}$)} &
  \multicolumn{1}{c|}{W(MgII)} &
  \multicolumn{1}{c|}{F(OII)} &
  \multicolumn{1}{c|}{F(OII$_{cont}$)} &
  \multicolumn{1}{c|}{W(OII)} &
  \multicolumn{1}{c|}{F(OIII)} &
  \multicolumn{1}{c|}{F(OIII$_{cont}$)} &
  \multicolumn{1}{c|}{W(OIII)} \\
\hline
  J031240 & 0.891 & 53.45(2.04) & 0.55 & 51.649 & $<$0.75(0.22) & 0.31 & $<$1.266 & --- & --- & ---\\
  J081445 & 1.17 & 28.35(0.81) & 0.34 & 38.378 & 1.05(0.21) & 0.23 & 2.129 & --- & --- & ---\\
  J081916 & 0.42568 & 54.32(1.90) & 0.44 & 87.424 & 2.11(0.51) & 0.32 & 4.669 & 18.98(0.21) & 0.23 & 57.52\\
  J083544 & 1.3267 & 35.90(0.60) & 0.42 & 37.086 & 1.03(0.34) & 0.23 & 1.895 & --- & --- & ---\\
  J084305 & 0.894 & 29.23(1.12) & 0.4 & 38.984 & 0.81(0.12) & 0.26 & 1.67 & --- & --- & ---\\
  J085220 & 0.8542 & 55.74(1.24) & 0.62 & 48.37 & $<$0.57(0.09) & 0.39 & $<$0.798 & --- & --- & ---\\
  J085759 & 0.7458 & 38.15(1.75) & 0.52 & 42.221 & $<$0.70(0.20) & 0.30 & $<$1.326 & 2.72(0.24) & 0.19 & 8.39\\
  J090244 & 0.4365 & 76.41(1.24) & 0.41 & 129.767 & 1.96(0.32) & 0.26 & 5.217 & 8.30(0.12) & 0.22 & 25.71\\
  J090514 & 1.2895 & 30.33(0.92) & 0.32 & 41.933 & --- & --- & --- & --- & --- & ---\\
  J092635 & 0.46499 & 21.46(1.47) & 0.48 & 30.273 & $<$0.93(0.26) & 0.40 & $<$1.589 & 2.34(0.20) & 0.34 & 4.76\\
  J094309 & 1.2691 & 59.33(0.80) & 0.76 & 34.26 & $<$1.02(0.27) & 0.48 & $<$0.937 & --- & --- & ---\\
  J094511 & 0.7578 & 24.72(0.97) & 0.21 & 66.531 & 1.04(0.12) & 0.14 & 4.369 & 1.68(0.15) & 0.09 & 10.45\\
  J103837 & 0.61978 & 46.44(1.50) & 0.6 & 47.799 & 0.83(0.19) & 0.36 & 1.441 & 7.60(0.20) & 0.18 & 25.4\\
  J104556 & 0.995 & 45.21(0.72) & 0.75 & 30.127 & 1.87(0.13) & 0.47 & 2.017 & --- & --- & ---\\
  J105502 & 0.41657 & 105.10(2.74) & 1.57 & 47.211 & 3.81(0.66) & 1.00 & 2.697 & 15.22(0.25) & 0.61 & 17.65\\
  J110805 & 0.536 & 48.87(2.24) & 0.67 & 47.246 & $<$0.90(0.27) & 0.37 & $<$1.609 & --- & --- & ---\\
  J124728 & 0.45396 & 103.90(3.93) & 1.46 & 48.829 & $<$2.16(0.65) & 0.87 & $<$1.714 & $<$1.66(0.53) & 0.61 & $<$1.88\\
  J133004 & 0.3574 & 77.34(1.63) & 1.48 & 38.477 & 14.17(0.41) & 1.14 & 9.159 & 32.27(0.19) & 0.62 & 38.14\\
  J135846 & 0.845 & 30.23(1.17) & 0.59 & 27.89 & --- & --- & --- & --- & --- & ---\\
  J141056 & 0.6743 & 70.61(1.66) & 0.82 & 51.52 & 3.17(0.22) & 0.55 & 3.463 & 9.66(0.24) & 0.33 & 17.34\\
  J142232 & 1.079 & 20.47(0.66) & 0.35 & 27.924 & $<$0.69(0.22) & 0.24 & $<$1.381 & --- & --- & ---\\
  J142902 & 0.4393 & 63.20(1.14) & 0.6 & 72.932 & $<$0.95(0.27) & 0.41 & $<$1.612 & 9.96(0.15) & 0.32 & 21.48\\
  J143531 & 1.5569 & 23.77(0.52) & 0.14 & 65.202 & $<$0.90(0.41) & 0.08 & $<$4.445 & --- & --- & ---\\
  J150042 & 0.7523 & 26.38(1.18) & 0.27 & 55.539 & 2.86(0.16) & 0.18 & 8.927 & 9.93(0.40) & 0.12 & 48.96\\
  J150210 & 0.6297 & 75.86(3.05) & 1.16 & 39.972 & $<$1.21(0.34) & 0.54 & $<$1.379 & 1.00(0.22) & 0.29 & 2.13\\
  J151201 & 0.933 & 30.03(1.47) & 0.43 & 36.746 & $<$0.64(0.19) & 0.25 & $<$1.304 & --- & --- & ---\\
  J151944 & 0.5339 & 70.00(1.95) & 1.14 & 40.113 & $<$1.22(0.22) & 0.69 & $<$1.059 & 2.05(0.15) & 0.42 & 3.15\\
  J154445 & 0.5478 & 72.65(5.25) & 1.17 & 39.974 & 4.14(0.47) & 0.68 & 3.904 & 10.41(0.39) & 0.36 & 18.76\\
  J155427 & 0.5718 & 59.26(1.45) & 1.14 & 33.198 & 1.19(0.15) & 0.66 & 1.15 & 5.24(0.17) & 0.34 & 9.74\\
  J160329 & 1.4124 & 18.79(0.86) & 0.33 & 23.43 & $<$1.48(0.64) & 0.21 & $<$2.981 & --- & --- & ---\\
  J160332 & 1.0439 & 49.63(0.72) & 0.27 & 89.368 & --- & --- & --- & --- & --- & ---\\
  J161022 & 1.986 & 21.17(0.71) & 0.32 & 22.206 & --- & --- & --- & --- & --- & ---\\
  J170845 & 0.5855 & 58.41(1.20) & 0.68 & 54.199 & $<$0.74(0.16) & 0.41 & 1.137 & 5.41(0.16) & 0.30 & 11.49\\
  J175610 & 0.6762 & 48.82(1.34) & 0.79 & 37.08 & 5.02(0.17) & 0.45 & 6.697 & 7.70(0.19) & 0.24 & 19.16\\
  J223210 & 0.27605 & 278.50(5.16) & 3.6 & 60.247 & $<$3.43(0.59) & 2.26 & $<$1.19 & 20.49(0.29) & 1.14 & 14.13\\
  J233237 & 1.471 & 27.12(0.48) & 0.27 & 40.913 & $<$0.60(0.22) & 0.16 & $<$1.505 & --- & --- & ---\\
  J234953 & 1.2779 & 28.73(0.56) & 0.29 & 42.909 & 1.14(0.234) & 0.19 & 2.676 & --- & --- & ---\\
\hline\end{tabular}

\end{table*}


\section{Complete light curves}

Here we provide the complete light curves for all objects; first the three year light curves for all sample objects, and then the ten year light curves for sixteen objects detected as CRTS transients.


\begin{figure*}
\centering
\includegraphics[width=1.0\textwidth,angle=0]{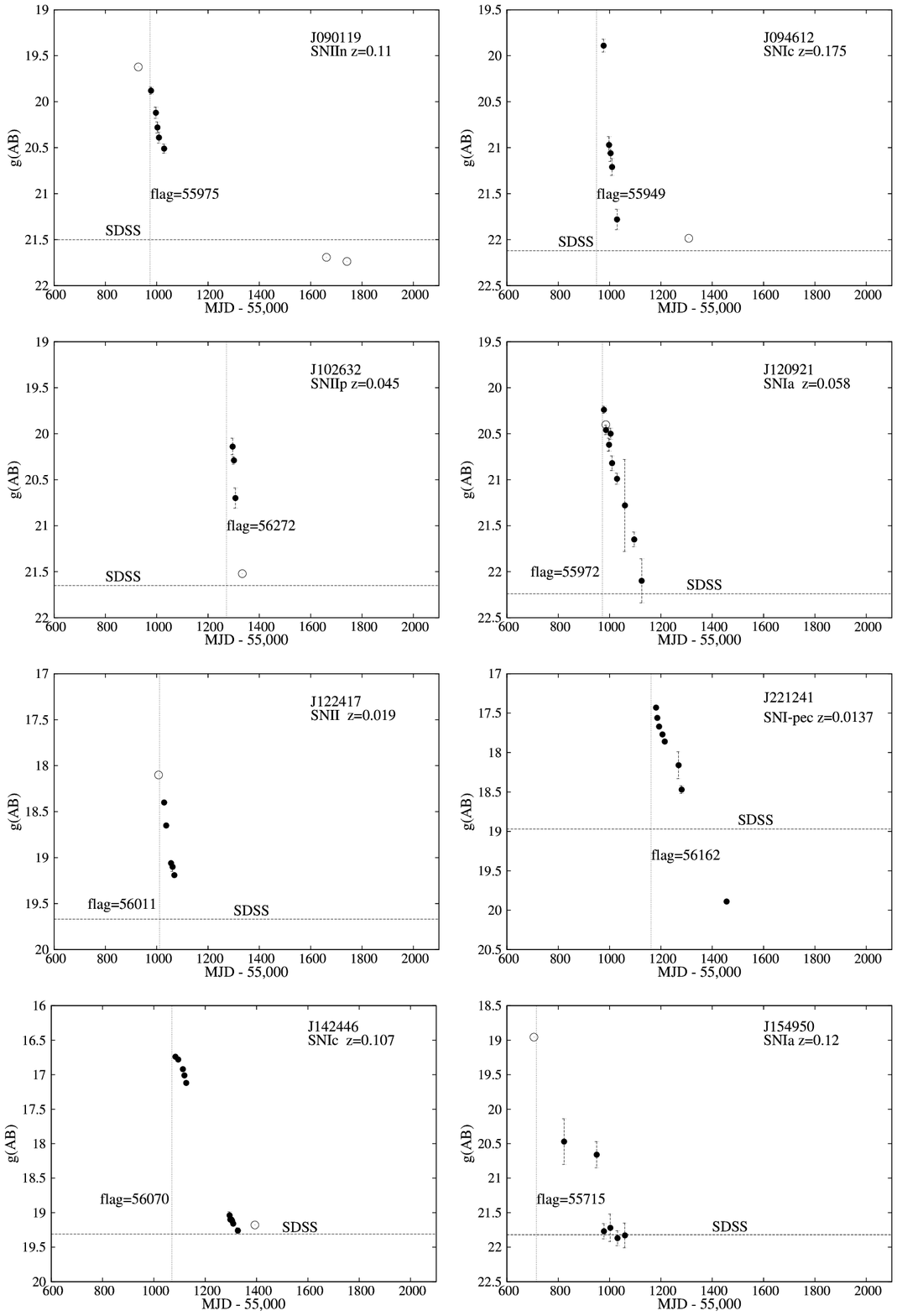}

\vspace{-3.0cm}
\caption{\it\small Three year light curves in $g$-band. (a) Objects which are known to be SNe. Solid symbols are LT data points; open symbols are PS1 data points. The vertical dotted line shows the date when flagged as a transient by PS1; the horizontal dotted line indicates the SDSS g-magnitude, approximately a decade earlier.}
\label{fig:lcs}
\end{figure*}

\begin{figure*}
\centering
\includegraphics[width=1.0\textwidth,angle=0]{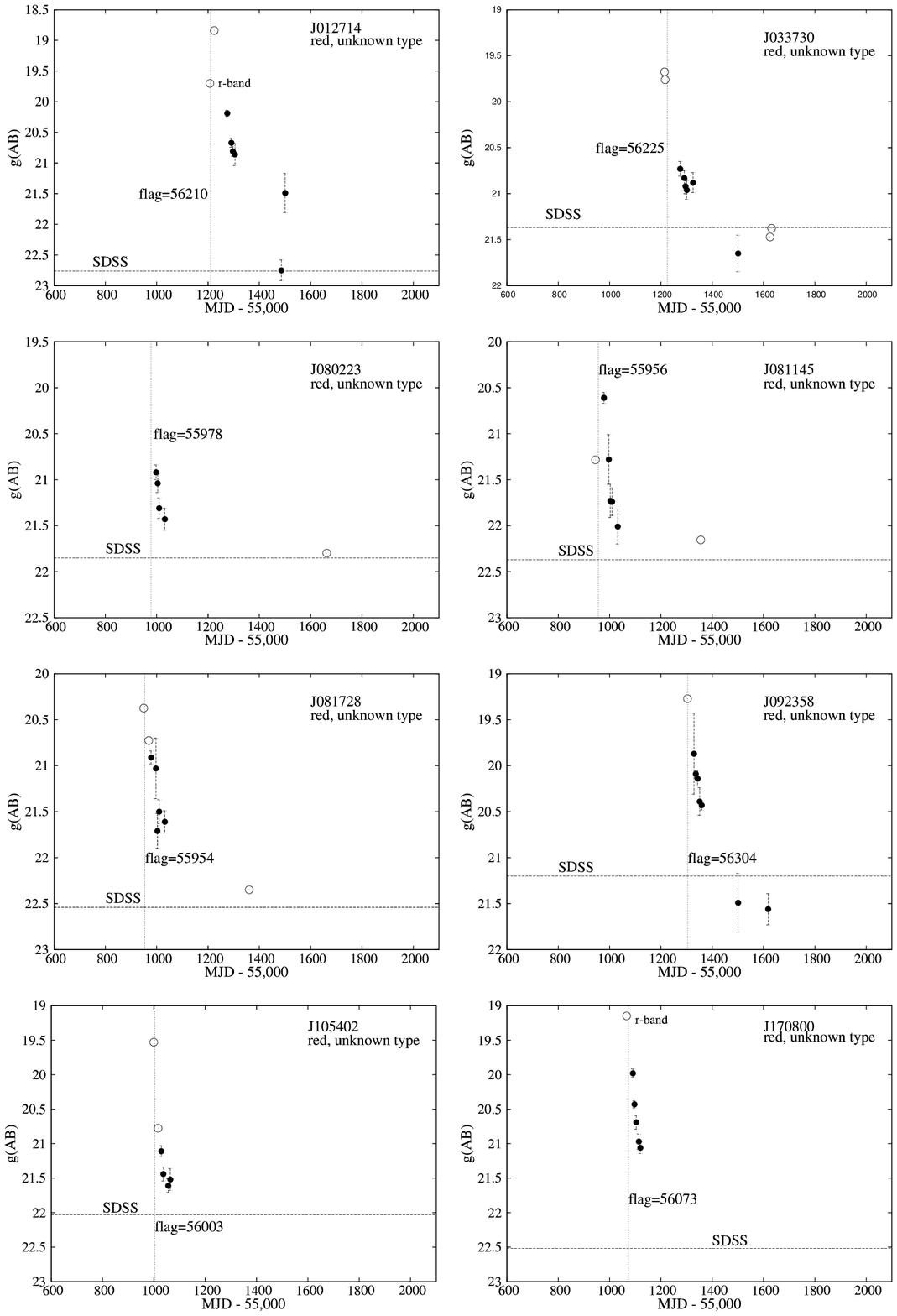}

\vspace{-3.0cm}
\caption{\it\small Three year light curves in $g$-band. (b) Objects likely to be SNe. Symbols as in Fig. B1}
\end{figure*}

\begin{figure*}
\centering
\includegraphics[width=1.0\textwidth,angle=0]{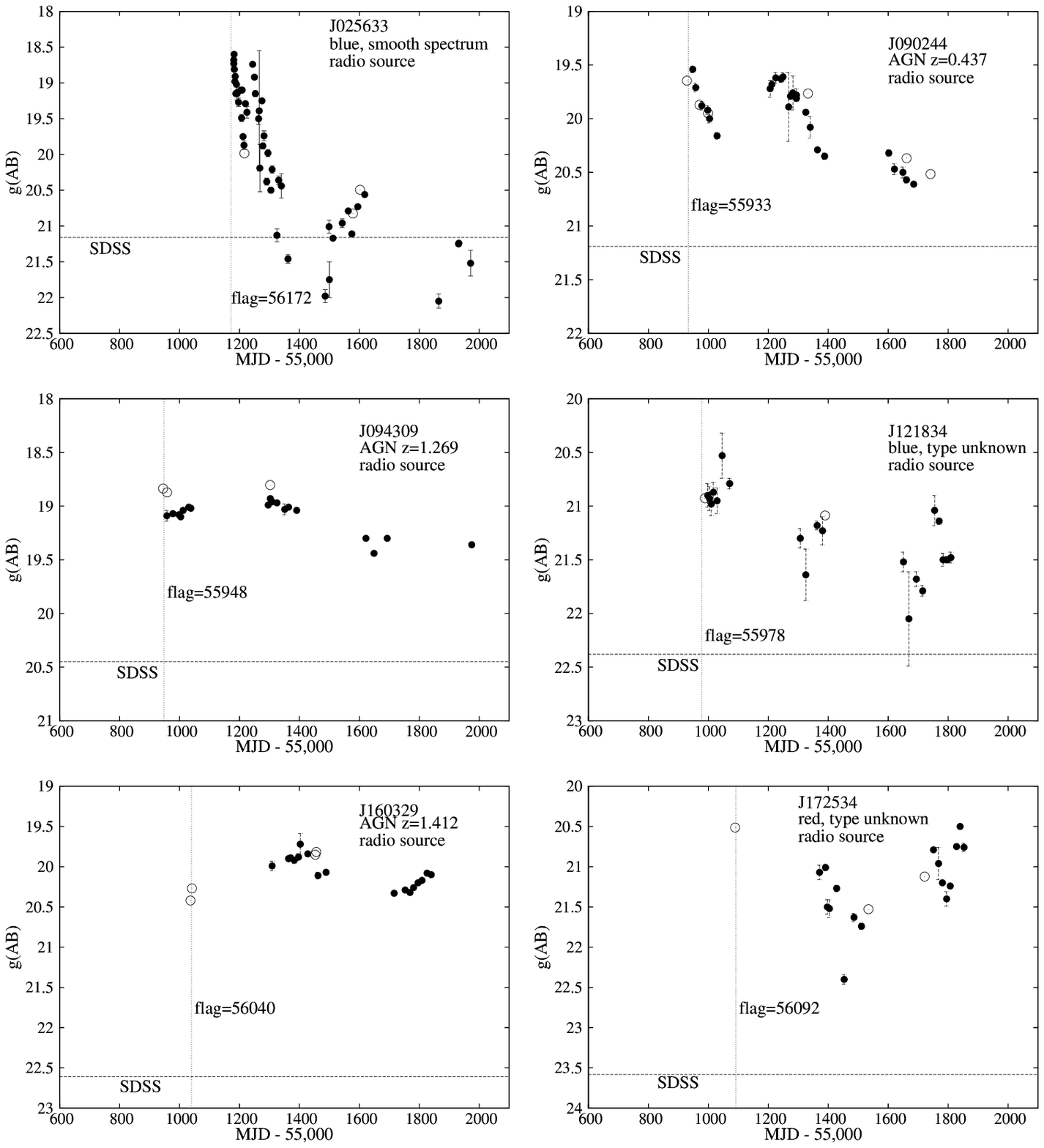}

\vspace{-3.0cm}
\caption{\it\small Three year light curves in $g$-band. (c) Objects which are radio sources; most but not all also known to be AGN. Symbols as in Fig. B1}
\end{figure*}

\begin{figure*}
\centering
\includegraphics[width=1.0\textwidth,angle=0]{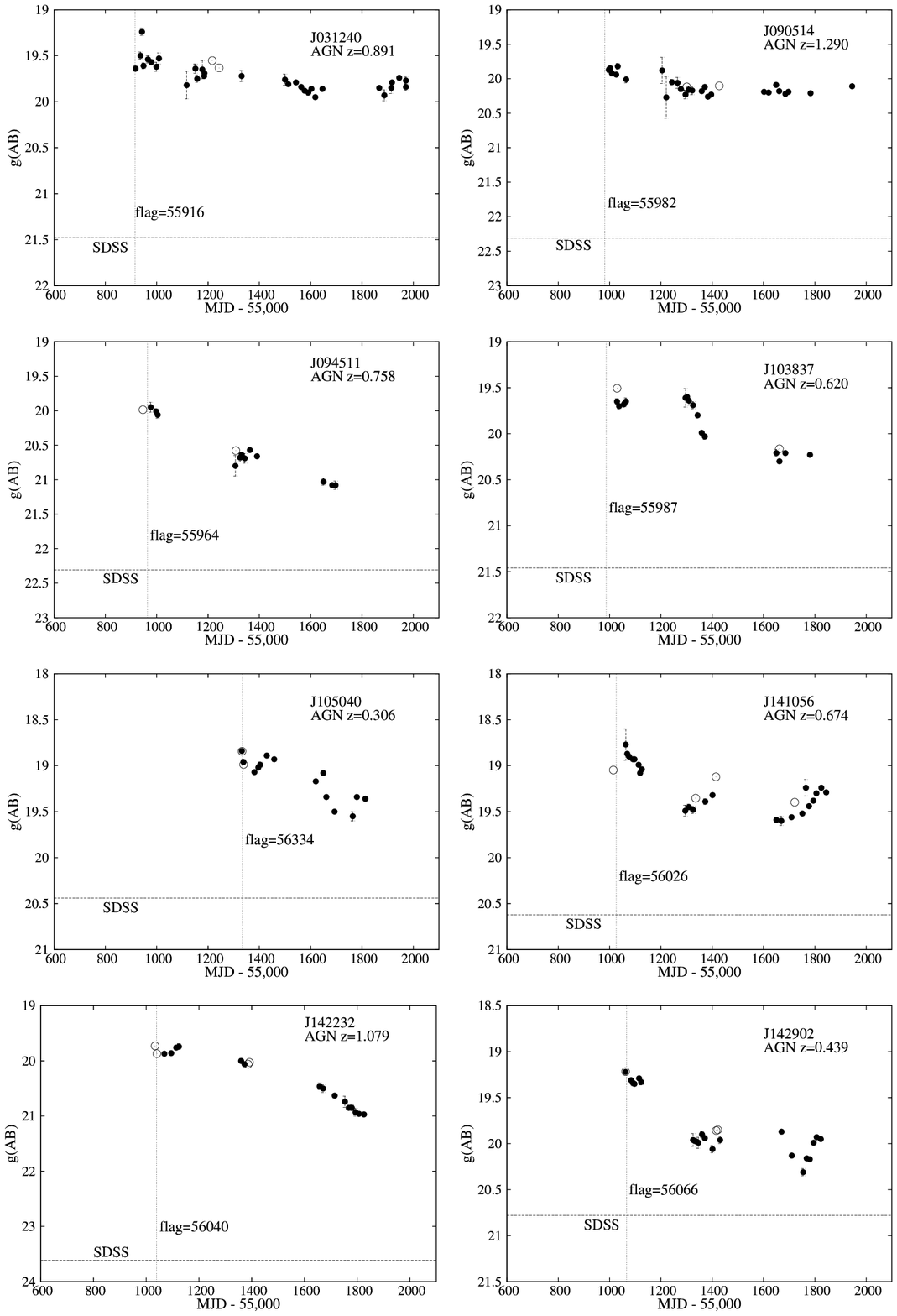}

\vspace{-3.0cm}
\caption{\it\small Three year light curves in $g$-band. (d) Objects which are AGN and have been falling since being flagged by PS1 - first eight of sixteen. Symbols as in Fig. B1}
\end{figure*}

\begin{figure*}
\centering
\includegraphics[width=1.0\textwidth,angle=0]{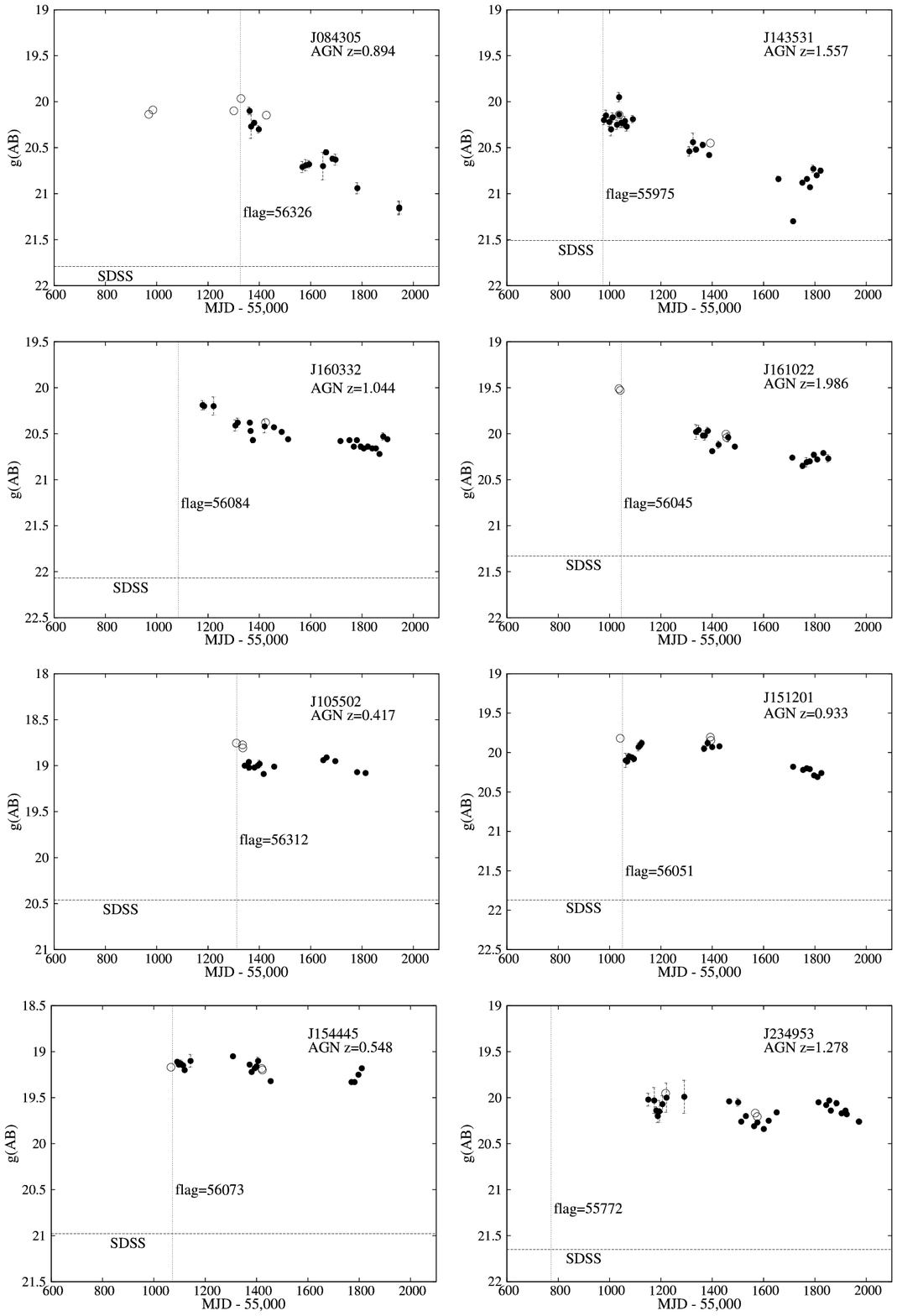}

\vspace{-3.0cm}
\caption{\it\small Three year light curves in $g$-band. (e) Objects which are AGN and have been falling since being flagged by PS1 - second eight of sixteen. Symbols as in Fig. B1}
\end{figure*}

\begin{figure*}
\centering
\includegraphics[width=1.0\textwidth,angle=0]{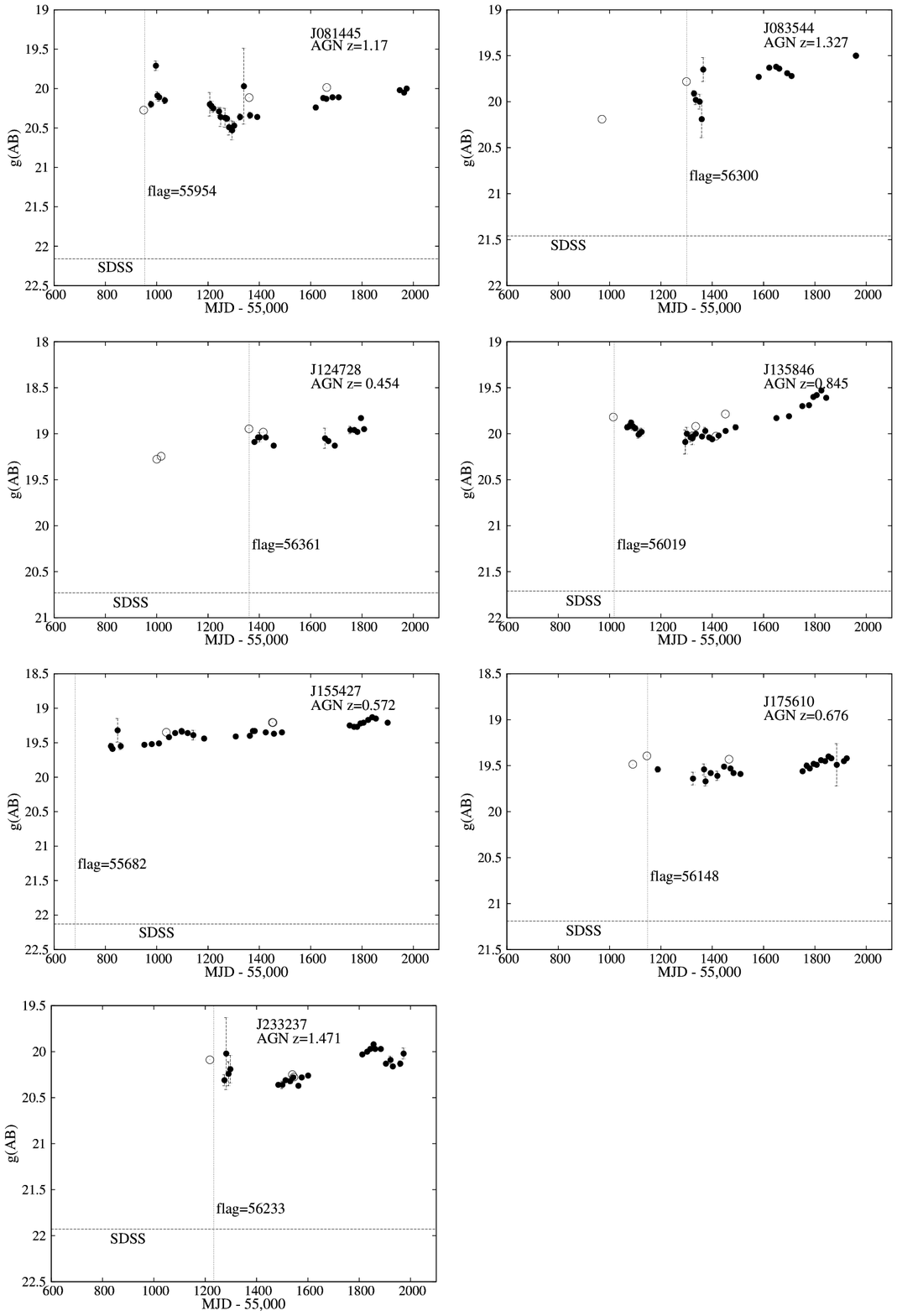}

\vspace{-3.0cm}
\caption{\it\small Three year light curves in $g$-band. (f) Objects which are AGN and have been rising since being flagged by PS1. Symbols as in Fig. B1}
\end{figure*}

\begin{figure*}
\centering
\includegraphics[width=1.0\textwidth,angle=0]{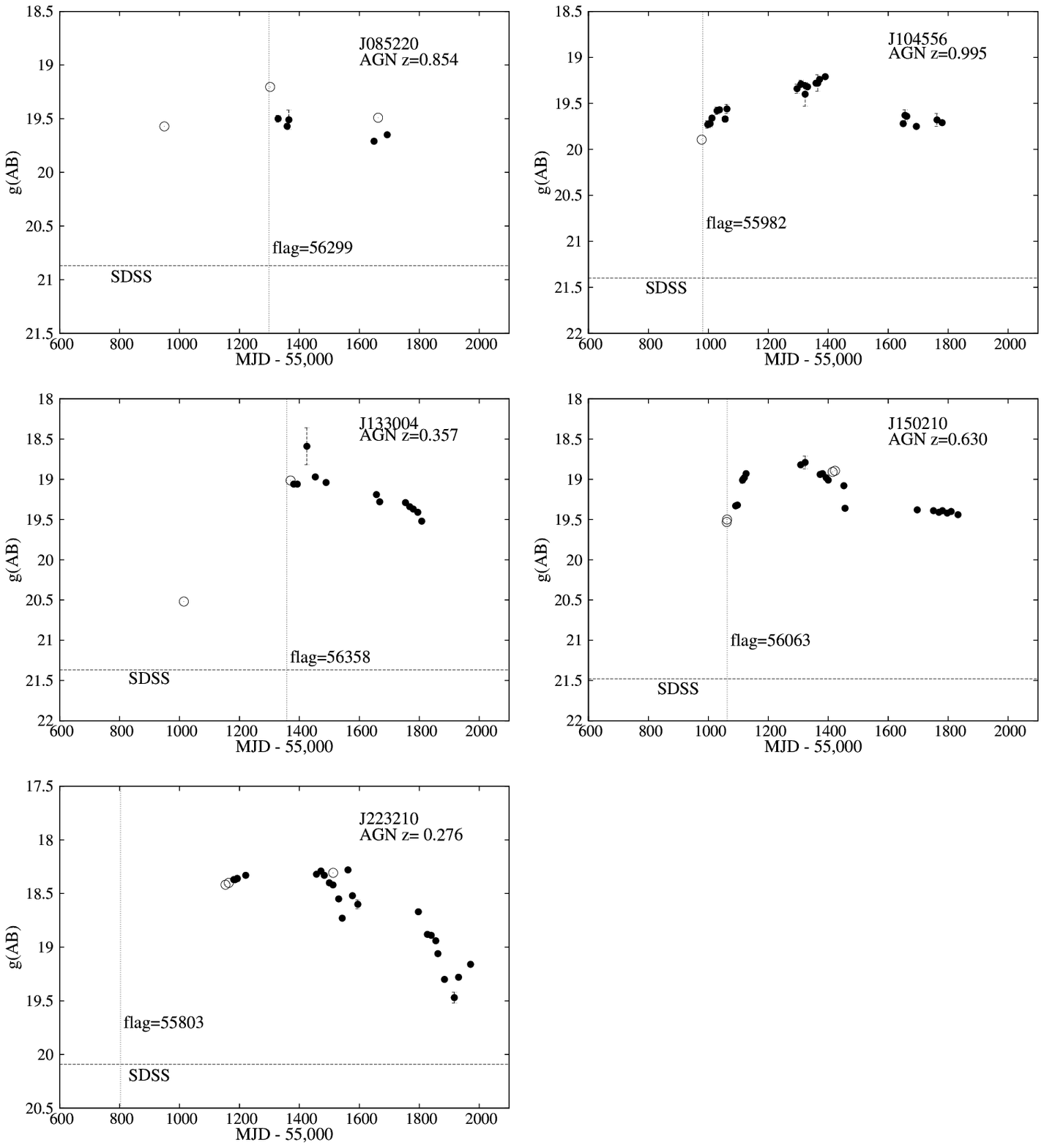}

\vspace{-3.0cm}
\caption{\it\small Three year light curves in $g$-band. (g) Objects which are AGN and have peaked during our monitoring period. Symbols as in Fig. B1}
\end{figure*}

\begin{figure*}
\centering
\includegraphics[width=1.0\textwidth,angle=0]{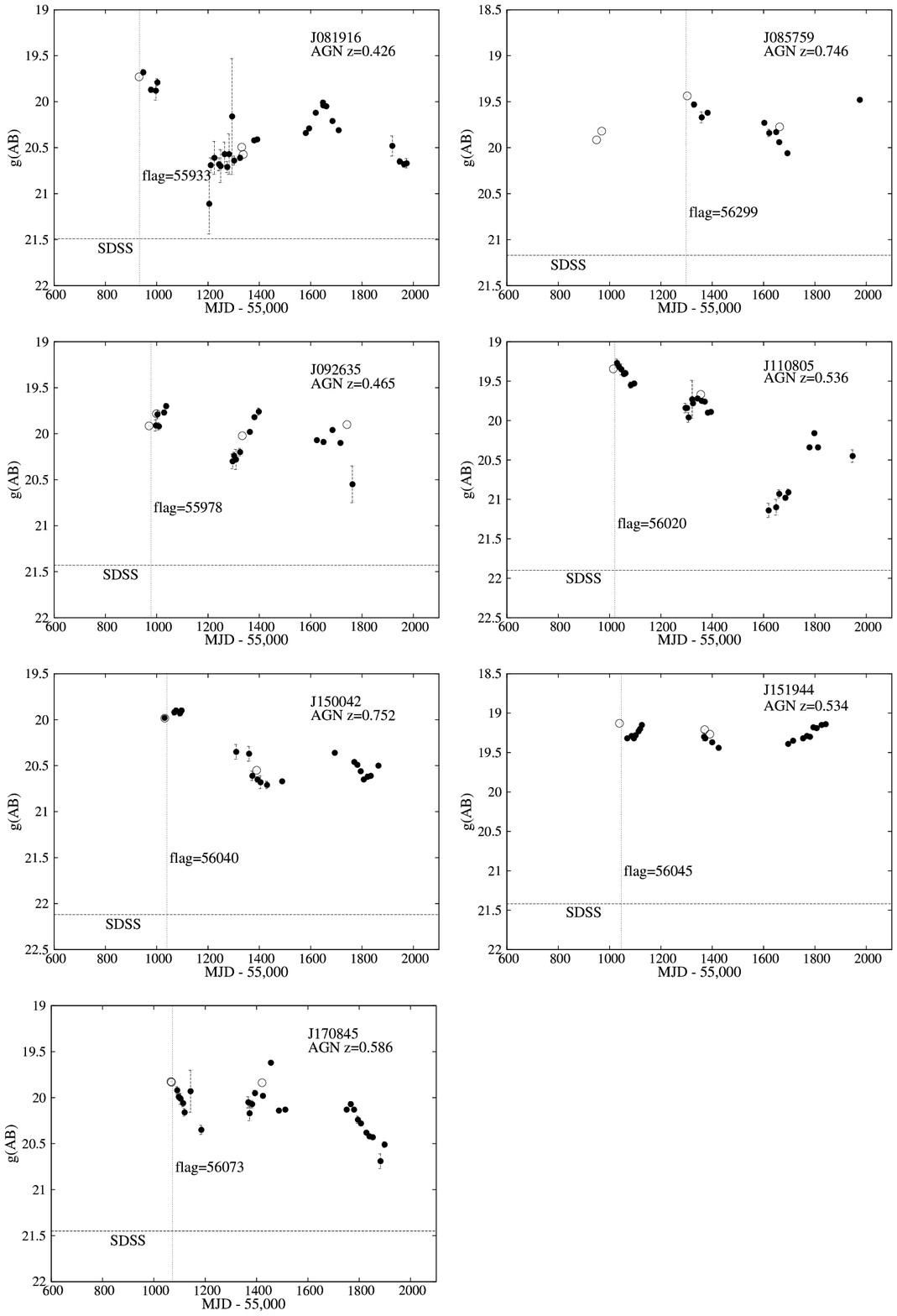}

\vspace{-3.0cm}
\caption{\it\small Three year light curves in $g$-band. (h) Objects which are AGN and show complex light curves, i.e. not simply falling or rising. Symbols as in Fig. B1}
\end{figure*}

\begin{figure*}
\centering
\includegraphics[width=1.0\textwidth,angle=0]{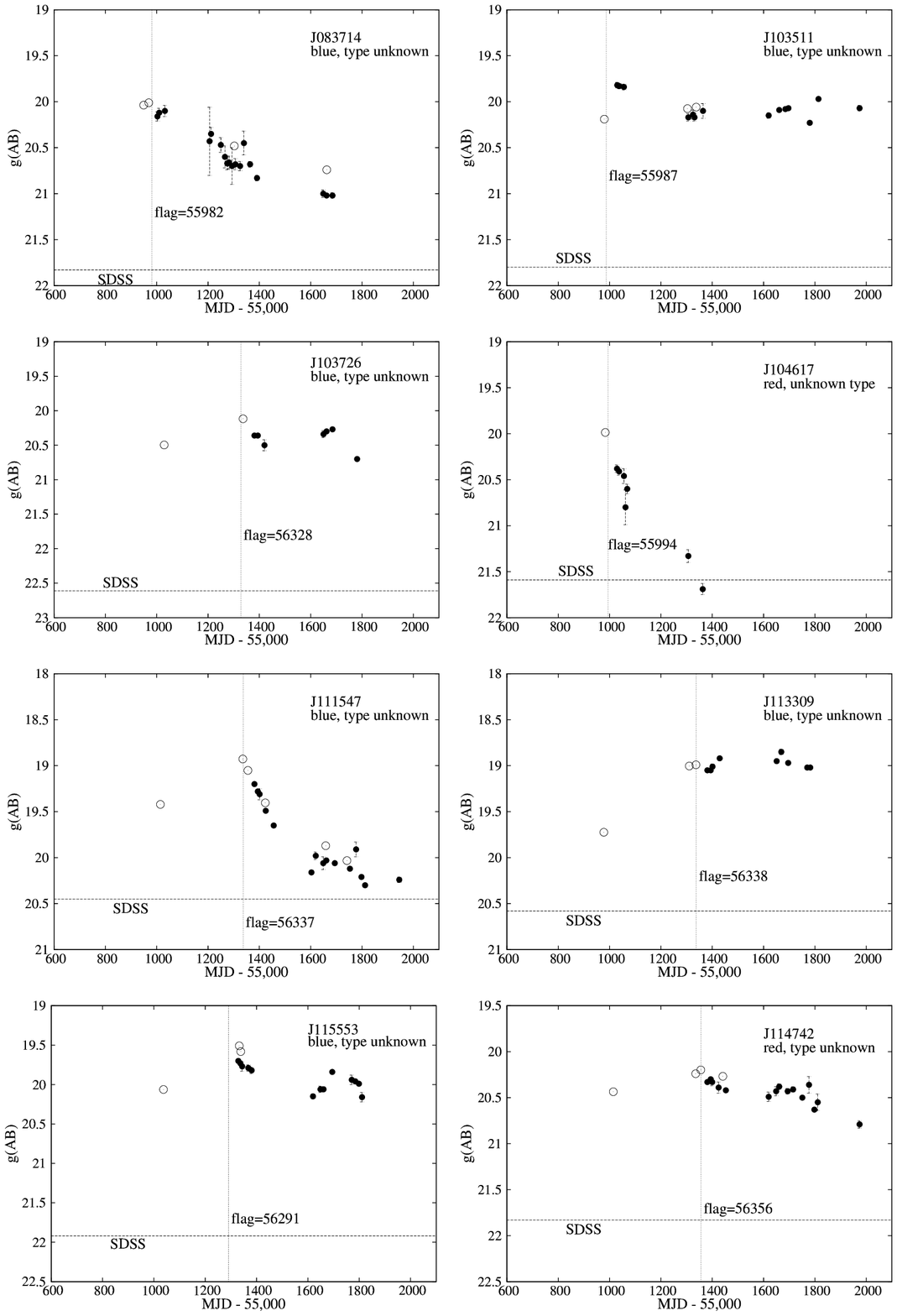}

\vspace{-3.0cm}
\caption{\it\small Three year light curves in $g$-band. (i) Objects which are not known to be AGN, but likely are, based on similarity in colour and light curve shape. Symbols as in Fig. B1. (Note added in revised version: since defining the sample and writing the paper we have now in fact obtained spectra for all these objects apart from J104617, and all these did indeed turn out to be AGN).}
\end{figure*}

\clearpage
\begin{figure*}
\centering
\includegraphics[width=1.0\textwidth,angle=0]{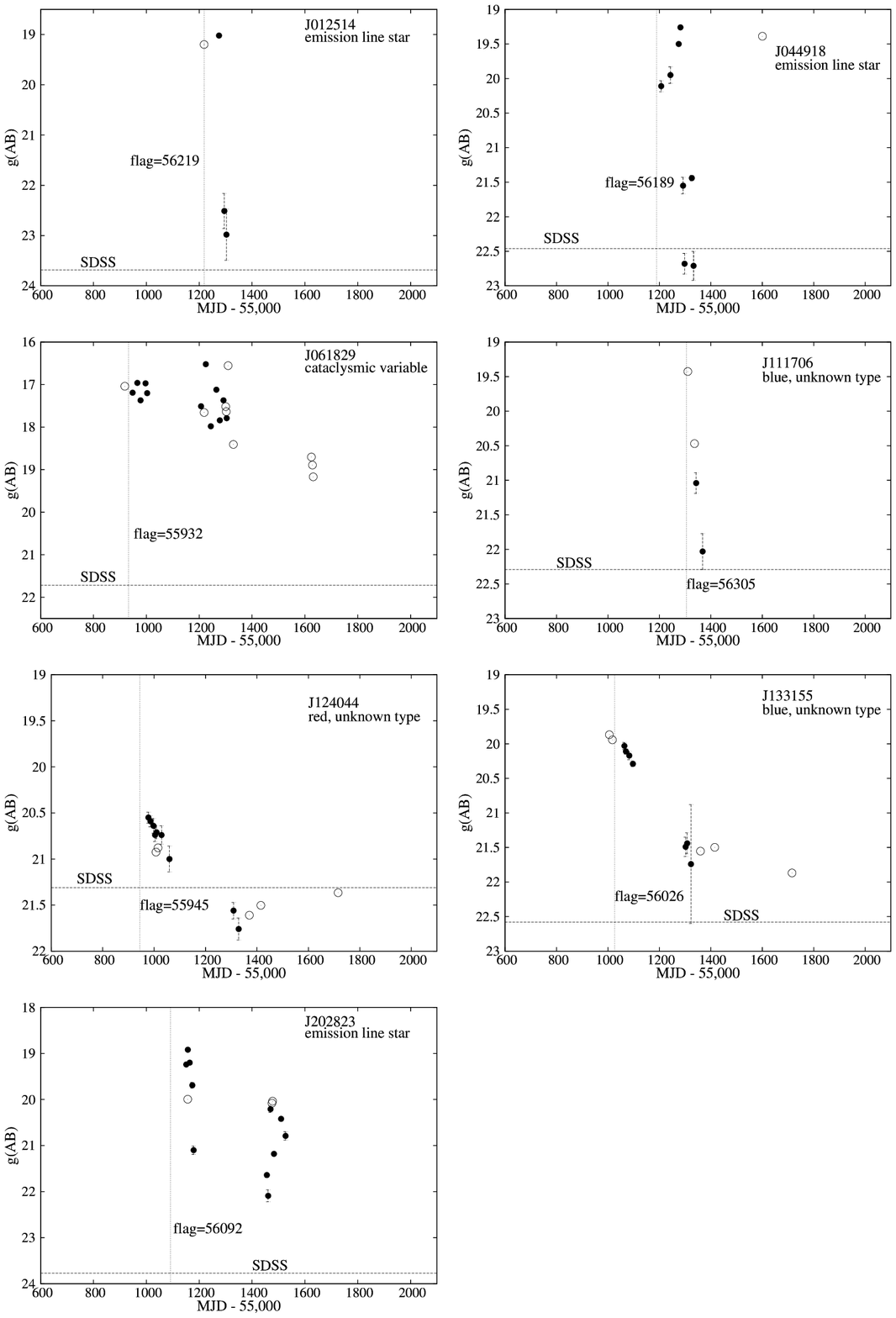}

\vspace{-3.0cm}
\caption{\it\small Three year light curves in $g$-band. (i) Other objects - four emission line stars and four objects of unknown type. Symbols as in Fig. B1.}
\end{figure*}


\begin{figure*}
\centering
\includegraphics[width=1.0\textwidth,angle=0]{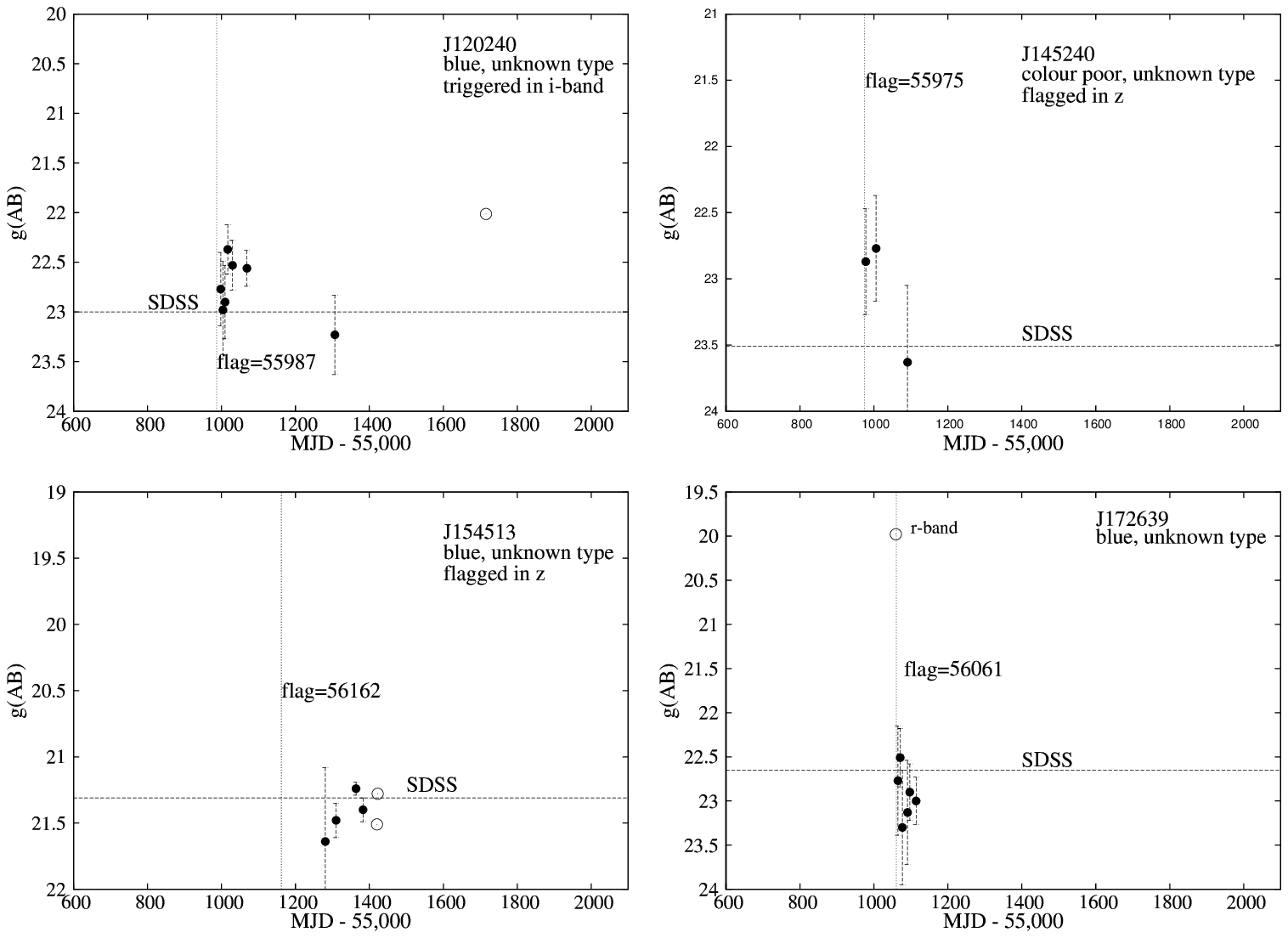}

\vspace{-3.0cm}
\caption{\it\small Three year light curves in $g$-band. (k) Objects where the data are too poor to come to a decision. Symbols as in Fig. B1}
\end{figure*}


\begin{figure*}
\centering
\includegraphics[width=1.0\textwidth,angle=0]{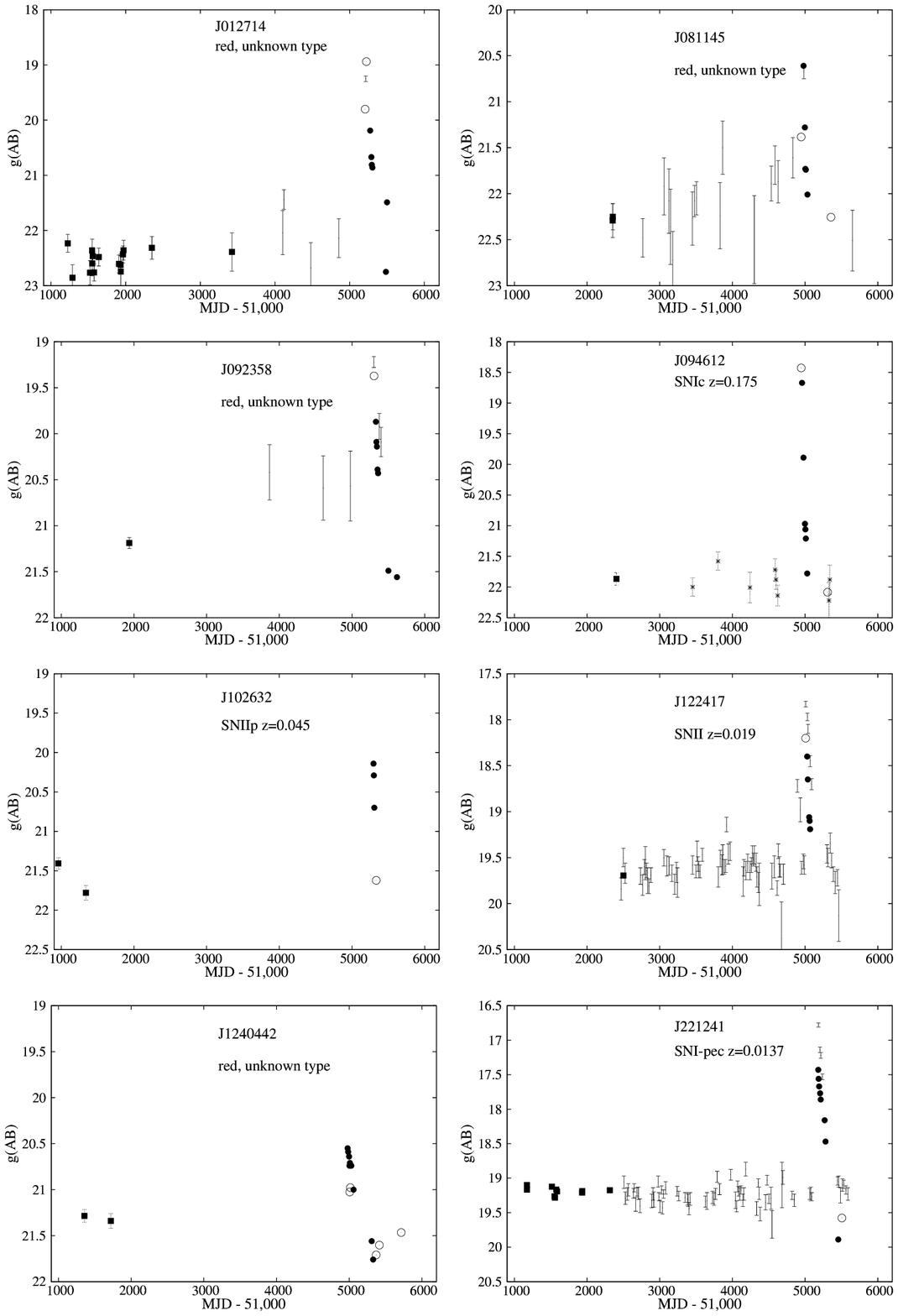}

\vspace{-3.0cm}
\caption{\it\small Long term light curves in $g$-band, for objects also detected as transients by CRTS. (a) Objects known or uspected to be SNe. Symbols as in Fig. B1}
\label{fig:lc-ten}
\end{figure*}


\begin{figure*}
\centering
\includegraphics[width=1.0\textwidth,angle=0]{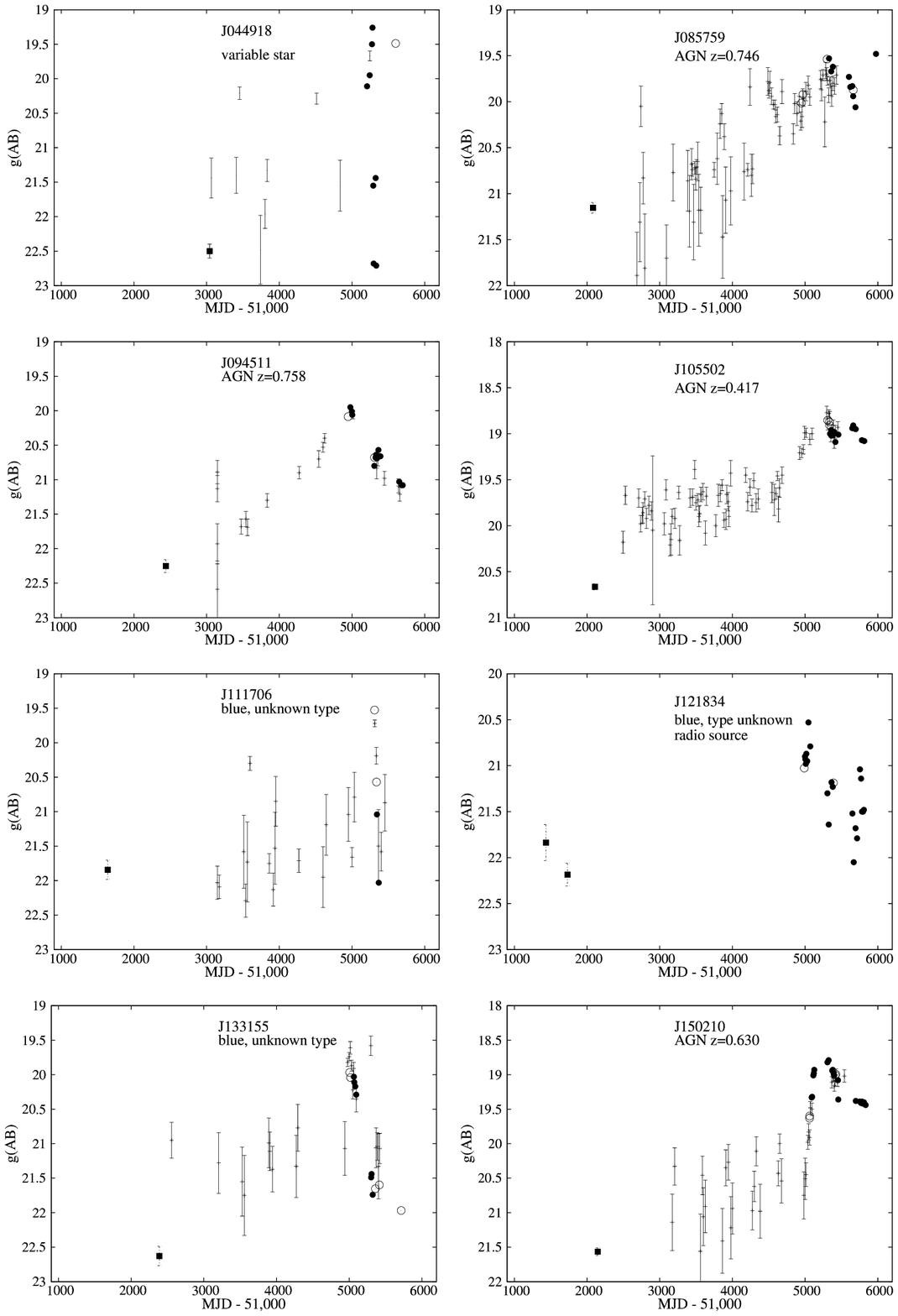}

\vspace{-3.0cm}
\caption{\it\small Long term light curves in $g$-band, for objects also detected as transients by CRTS. (a) Objects not known to be SNe. Symbols as in Fig. B1}
\end{figure*}


\label{lastpage}

\end{document}